\documentclass[
superscriptaddress,
bibnotes,
 amsmath,amssymb,
 aps,
 prb,twocolumn, 
longbibliography,
eprint,
floatfix,
reprint
]{revtex4-2}

\usepackage{amsmath}
\usepackage{graphicx}
\usepackage{dcolumn}
\usepackage{bm}
\usepackage{hyperref}
\usepackage{braket}
\usepackage{framed}
\usepackage{bbding}
\usepackage[dvipsnames]{xcolor}
\usepackage{array}
\usepackage{floatrow}
\usepackage{bbding}

\newcommand*\colourcheck[1]{%
  \expandafter\newcommand\csname #1check\endcsname{\textcolor{#1}{\ding{52}}}%
}

\definecolor{darkred}{rgb}{0.66, 0, 0}
\definecolor{lblue}{rgb}{0.33, 0.33, 1}
\definecolor{lcyan}{rgb}{0, 0.8, 0.8}
\definecolor{darkgreen}{rgb}{0, 0.5, 0}

\newcommand{\redbox}[2][0 pt]{\raisebox{#1}{\fcolorbox{black}{darkred}{\rule{0pt}{2pt}\rule{2pt}{0pt}}}}

\newcommand{\orangebox}[2][0 pt]{\raisebox{#1}{\fcolorbox{black}{orange}{\rule{0pt}{2pt}\rule{2pt}{0pt}}}}

\newcommand{\brownbox}[2][0 pt]{\raisebox{#1}{\fcolorbox{black}{brown}{\rule{0pt}{2pt}\rule{2pt}{0pt}}}}

\newcommand{\bluebox}[2][0 pt]{\raisebox{#1}{\fcolorbox{black}{lblue}{\rule{0pt}{2pt}\rule{2pt}{0pt}}}}

\newcommand{\cyanbox}[2][0 pt]{\raisebox{#1}{\fcolorbox{black}{lcyan}{\rule{0pt}{2pt}\rule{2pt}{0pt}}}}

\newcommand{\f}{\frac}
\def\({\left(}
\def\){\right)}

\newcommand{\lb}{\left[}
\newcommand{\rb}{\right]}
\newcommand{\half}{\frac{1}{2}}

\newcommand{\beq}{\begin{equation}}
\newcommand{\eeq}{\end{equation}}
\newcommand{\beqs}{\begin{equation*}}
\newcommand{\eeqs}{\end{equation*}}
\newcommand{\beqar}{\begin{eqnarray}}
\newcommand{\eeqar}{\end{eqnarray}}
\newcommand{\bal}{\begin{aligned}}
\newcommand{\eal}{\end{aligned}}
\newcommand{\bmult}{\begin{multline}}
\newcommand{\emult}{\end{multline}}

\newcommand{\bfa}{{\bf a}}

\newcommand{\bfj}{{\bf j}}
\newcommand{\bfk}{{\bf k}}
\newcommand{\bfp}{{\bf p}}
\newcommand{\bfq}{{\bf q}}
\newcommand{\bfr}{{\bf r}}
\newcommand{\bfdd}{{\bf d}}
\newcommand{\bfd}{{\bm \Delta}}
\newcommand{\bfg}{{\boldsymbol \gamma}}
\newcommand{\bfs}{{\boldsymbol \sigma}}
\newcommand{\bfG}{{\bf G}}
\newcommand{\bft}{{\bf t}}
\newcommand{\bfR}{{\bf R}}
\newcommand{\bfK}{{\bf K}}
\newcommand{\bfS}{{\bf S}}
\newcommand{\bfQ}{{\bf Q}}

\def\a{\alpha}
\def\b{\beta}

\def\d{\delta}

\def\l{\lambda}

\def\s{\sigma}

\def\k{\kappa}

\def\w{\omega}

\begin{document}

\preprint{APS/123-QED}

\title{Doped moir\'e magnets: renormalized flat bands and excitonic phases}

\author{Ilia Komissarov}
 \email{i.komissarov@columbia.edu}
\affiliation{Department of Physics, Columbia University, New York, NY 10027, USA}%

\author{Onur Erten}
 \email{onur.erten@asu.edu}
\affiliation{Department of Physics, Arizona State University, Tempe, AZ - 85287, USA}

\author{Pouyan Ghaemi}
 \email{pghaemi@ccny.cuny.edu}
\affiliation{Physics Department, City College of the City University of New York, NY 10031, USA}
\affiliation{Physics Program, Graduate Center of City University of New York, NY 10031, U.S.A.}

\begin{abstract}
We explore the phase diagram of a twisted bilayer of strongly interacting electrons on a honeycomb lattice close to half-filling using the slave boson mean-field theory. Our analysis indicates that a variety of new phases can be realized as a function of chemical doping and twist angle. In particular, we find a non-magnetic excitonic insulating phase that breaks the translational symmetry of the underlying moir\'e pattern. This phase results from the interplay of strong Coulomb interactions and the twist angle. In addition, we show that the features of the renormalized dispersion such as the magic angles depend significantly on the interactions. Our results highlight the rich physics arising in doped moir\'e superlattices of Mott insulators.  
\end{abstract}

\maketitle



\section{Introduction}
The 
discovery of superconductivity in twisted bilayer graphene \cite{Cao2018}
has sparked extensive research into moir\'e superlattices. The quantum interference of electronic wavefunctions in twisted or misaligned bilayers leads to the suppression of the kinetic energy of electrons relative to the electron-electron interaction energy. Leveraging the high tunability of such platforms, much effort is made to explore and control correlated electronic phases of matter \cite{Jorio2022,Wang2020,Seiler2022}, which have long been a complex and intriguing area of research \cite{Paschen2021}. Correlated phases of matter can also be realized in bulk $d$- and $f$-electron materials such as cuprates \cite{lee2004} or heavy fermions \cite{Coleman_JPCM2001}. In these systems, the strong electron-electron interaction surpasses the kinetic energy due to the absence of screening. The correlated electronic states realized via the twisting share many similarities with these phases, however, some important differences are notable \cite{Song_2022}.

\begin{figure}[t!]
\includegraphics[width=0.7\linewidth]{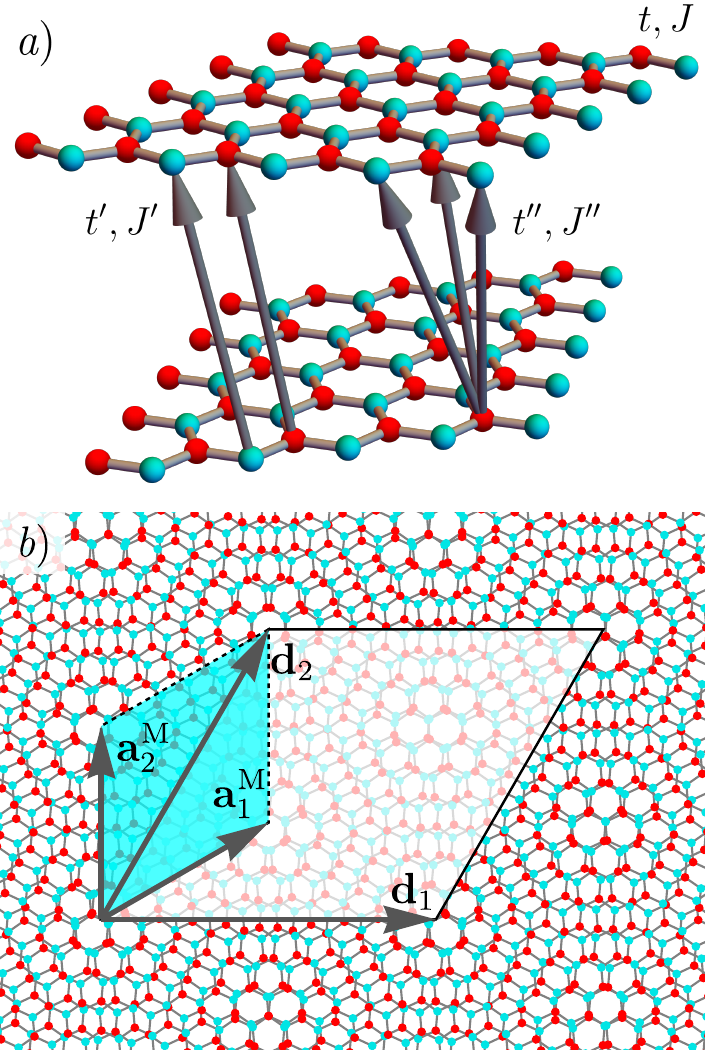}
\caption{$\textit{a})$ The depiction of a twisted bilayer doped Mott insulator with the intra- and interlayer hopping and superexchange parameters. The intralayer hopping and superexchange are denoted with $t$ and $J$. The interlayer hoppings (superexchange) terms considered are indicated with arrows: $J'$ and $t'$ correspond to AA and BB terms, and $J''$ and $t''$ interpolate between different sublattices of different layers. $\textit{b})$ At high values of the interlayer superexchange $J'$ and sufficient hole doping $\mu_0 < 0$, the ground state of the system is the excitonic insulator of spinons characterized by the non-zero values of the order parameter $\chi_{\rm AA(BB)}=\braket{\hat f_{\rm A(B)}^{(1)\dagger} \hat f_{\rm A(B)}^{(2)}}$. The area of the unit cell in this phase (shaded with white) is three times larger than the area of the hexagonal moir\'e unit cell (shown with blue), and the lattice vectors in the real space are promoted from $\bfa_1^{\rm M}$ and $\bfa_2^{\rm M}$ to $\bold d_1$ and $\bold d_2$.}
\label{plot1}
\end{figure}

While moir\'e engineering of electronic phases in weakly-correlated layers such as semiconductors and graphene has been extensively studied, the research on the moir\'e superlattices composed of Mott insulators and magnets is at its seminal stages. To date, a limited number of theoretical investigations on moir\'e superlattices of magnetic materials have been conducted \cite{Hejazi2020, Akram_PRB2021, Hejazi_PRB2021, Akram_NanoLett2021, Nica2023, Kim_NanoLett2023, Akram_NanoLett2024, Keskiner_NanoLett2024, Das_arXiv2024}. Several of these predicted phases have been demonstrated experimentally \cite{Xu_NatNano2022, Song_Science2021, Xie_NatPhys2023}. One prominent example of such a platform is the twisted cuprates where the Dirac spectrum emerges with the formation of the Bogoliubov quasiparticle excitations of the $d$-wave superconducting phases \cite{PhysRevB.105.064501,twcup2,twc3,Liu2023,PhysRevB.105.245127}.


In this article, we explore the venue of small twist angle bilayers of honeycomb materials with strong Coulomb repulsion \cite{interactingdr} described by the $t-J$ model \cite{PhysRevB.18.3453,PhysRevB.108.035144}. Our study is motivated by the observation that many two-dimensional Mott insulators with novel magnetic quantum phases possess a honeycomb lattice structure \cite{Blei_APR2021}. Contrary to the twisted non-interacting materials, where the non-trivial correlated states are stabilized in a narrow range of dopings close to Fermi energy \cite{zaletel}, in the system considered here, due to the strong interactions in each layer, electronic states in a wide range of energies participate in the formation of the ground state. In this aspect, our approach is essentially different from the commonly used approximation that utilizes the projection into the flat topological bands at charge neutrality. Here, we use the entire band structure of the interacting continuum model, and the obstruction to the Wannierization present in the low-lying Chern bands is avoided. This sacrifices some of the ultraviolet details of the band structure due to the truncation of the Brillouin zone but preserves the physical picture of the local moments.

Our results show that twisted doped Mott insulators can host new types of quantum phases; one example of which is an excitonic insulator of spinons. The emergence of this phase is crucially related to both strong correlations in each layer and the relative twist of the two layers, which produces the moir\'e minibands. Within our mean-field analysis, the gaps between the minibands are controlled by the interactions, which enables the stabilization of the non-magnetic insulating phases when the Fermi level lies in one of such gaps. This excitonic insulator originates from the finite momentum hybridization of the interlayer spinons and breaks the translation symmetry of the underlying moir\'e lattice while the intralayer order parameters stay homogeneous. 


The rest of the paper is organized as follows. In \autoref{sec2}, we describe the parton mean-field theory for the strongly interacting hexagonal bilayer. The \autoref{sec3} details the phase diagram of the slave-boson mean field theory. Some aspects peculiar to the twisted bilayers such as the magic angles and the twist-angle dependence of the phases are also discussed. We conclude with a summary of our results and an outlook in \autoref{sec4}.


\section{Model and Methods}
\label{sec2}

We consider a twisted honeycomb bilayer as depicted in \autoref{plot1} $a)$. Assuming the limit of strong Coulomb interactions and small doping, we use the following $t-J$ model which captures both inter and intra-layer interactions:
\beq
\begin{aligned}
 \hat H_{t-J}  = & - \sum_{\braket{i j}} t_{i j} \hat P \hat c^\dagger_{i \s} \hat c_{j \s} \hat P \\ & +   \sum_{\braket{ij}} J_{ij} (\hat \bfS_i \cdot \hat \bfS_j - \frac{1}{4} \hat n_i \hat n_j)  -  \mu_0 \sum_i \hat c^{\dagger}_{i \s}  \hat c_{i \s}\, .
\end{aligned}
\label{tj}
\eeq
Here, we define $t_{ij} \equiv t$ and $J_{ij} \equiv J $ for the indices $i, j$ belonging to the neighboring cites of the honeycomb lattice in the same layer. The interlayer hoppings and magnetic couplings have the forms $t_{ij} = t'(|i-j|)$ and $J_{ij} = J'(|i-j|)$ for $i,j$ belonging to the same sublattice in both layers whereas $t_{ij} = t''(|i-j|)$, $J_{ij} = J''(|i-j|)$ when $i,j$ belong to opposite sublattices in the two layers. We assume that $t',t'',J',J''$ are smooth functions of their argument which justifies using the interacting analog of the Bistritzer-Macdonald model \cite{Bistritzer_2011}.

The spin operators are expressed in terms of Abrikosov fermions as
$\hat \bfS_i = \half \sum_{\s\s'} \hat f^\dagger_{i \s} \bfs_{\s \s'} \hat f_{i \s'} \,$ 
where $\bfs$ is a vector of Pauli matrices and $\hat P$ denotes the projector into the subspace of singly occupied sites. We implement the single-occupancy constraint via slave-boson formalism \cite{Coleman_book}
\beq
\hat c_{i\sigma} = \hat b^\dagger_i \hat f_{i\sigma}\, ,
\eeq
which leads to the holonomic condition
\beq
\sum_\sigma \hat f^\dagger_{i\sigma} \hat f_{i\sigma}+\hat b^\dagger_i \hat b_i = 1 \, .
\label{constr}
\eeq
The relation above will be taken into account at the mean-field level by introducing into the Hamiltonian a corresponding term with a constant Lagrange multiplier $\lambda$. Within our mean-field analysis, we consider solutions with uniform holon occupation $\delta \equiv \braket{\hat b^\dagger_i \hat b_i}$. We further decouple the superexchange term in \eqref{tj} in the direct and exchange channels as shown in Appendix \ref{appa}. In this work, we do not consider the superconducting state and retain the susceptibilities of the form $\chi_{ij}=\braket{\hat f^\dagger_{i} \hat f_{j}}$. The direct terms can be expressed in terms of $\delta$ through the constraint equation \eqref{constr} (see  Appendices \ref{appa}, \ref{appb}). The resulting mean-field Hamiltonian can be written as
\beq
\hat H = \hat H_{\rm intra}+\hat H_{\rm inter}+ (\lambda - \mu_0) \sum_\bfk \hat f^\dagger_{\bfk} \hat f_{\bfk}  +C\, ,
\label{modelH}
\eeq
where 
\beq
\hat f_\bfk = (\hat f^{(1)}_{\bfk A} ~~ \hat f^{(1)}_{\bfk B} ~~ \hat f^{(2)}_{\bfk A} ~~ \hat f^{(2)}_{\bfk B}) \, ,
\eeq
and indices $(1)$, $(2)$ denote the layer. The intralayer term is as follows:
\beq
\(\hat H_{\rm intra}\)_{\bfQ\bfQ'} = \d_{\bfQ\bfQ'}  \sum_{\bfk} \hat f^\dagger_{\bfk} H_{\rm intra}(\bfk) \hat f_{\bfk} \, ,
\eeq
where $\bfQ, \bfQ'$ lie within the moir\'e reciprocal lattice spanned by vectors $\bfd_n$ (see \autoref{hexplots}). 
The contribution from the two valleys can be presented as follows:
\beq
H_{\rm intra}(\bfk) = - \( \d t + \f{\chi J}{4}\) \begin{pmatrix}
K_\bfk^{(1)} & 0  \\
 0  & K_\bfk^{(2)} \, ,
\end{pmatrix}
\eeq
with the rotated hopping matrices
\beq
 K_\bfk^{(l)} = v_F \lb R(\pm \theta/2) (\bfk+\bfQ)\cdot(\hat \s_x, \hat \s_y) \rb\, , ~~  v_F = \f{3}{2} a t \, ,
\eeq
where we set the lattice constant $a$ equal to unity. The interlayer Hamiltonian $\hat H_{\rm inter}$ is a sum of the hopping $\hat H_{\rm inter}^t$ and superexchange $\hat H_{\rm inter}^J$ components as introduced below:
\begin{equation}
    (\hat H_{\rm inter})^{t(J)}_{\bfQ \bfQ'}  = \sum_{\bfk \l n m} \hat f^\dagger_{\bfk} H_{{\rm inter},m\l}^{t(J)} \hat f_{\bfk} \d_{\bfQ - \bfQ',\, \l (\bfd_m-\bfq_{ n}^{t(J)})} \, ,
\end{equation}
where $\l=\pm 1$ represent the hopping from layer 2 to layer 1 and the inverse process correspondingly, $m$ and $n$ run between 1 and 3, and $\bfq_n^t = 0$ \cite{Bistritzer_PNAS2011}. 

Inter-layer super-exchange has a momentum dependence of the general form
\begin{equation}
\langle \hat{f}_{i \alpha}^{(1) \dagger} \hat{f}_{j \beta}^{(2)} \rangle = \chi_{\alpha \beta} \sum_{n=1}^{3} e^{i \mathbf{q}_n^J \cdot \mathbf{j}}
\label{chians}
\end{equation}
 We consider two possible mean-field solutions where $ \bfq_n^J = 0$, and $ \bfq_n^J = \bfd_n$: in the latter choice, the inter-layer hybridization momenta are modified by the momentum imparted by the spinon inter-layer order parameters. The part of the Hamiltonian which corresponds to inter-layer  couplings assumes the form
\beq
H_{{\rm inter}, m \l}^t = 
 \begin{pmatrix}
0 &  T^t_m \\
0  & 0
\end{pmatrix} \d_{\l,1} + \({\rm h. c.}\) \d_{\l,-1} \, ,
\eeq
\beq
H_{{\rm inter}, m \l}^J = 
 \begin{pmatrix}
0 &  T^J_m \\
0  & 0
\end{pmatrix} \d_{\l,1} + \({\rm h. c.}\) \d_{\l,-1} \, ,
\eeq

\begin{equation}
    T_m^t = - \begin{pmatrix}
         \d t'&  \w^{m-1} \d t'' \\
         \w^{-(m-1)} \d t''  & \d t' 
    \end{pmatrix} \, ,
\end{equation}
\begin{equation}
    T_m^J = - \begin{pmatrix}
        \f{\chi_{\rm AA}J' }{4}&  \w^{m-1}\f{\chi_{\rm AB}J''}{4} \\
        \w^{-(m-1)} \f{\chi_{\rm AB}J'' }{4} &  \f{\chi_{\rm BB}J' }{4} 
    \end{pmatrix} \, ,
\end{equation}
where $\omega = e^{2 \pi i/3}$.
The constant term  $C$ depends on the choice of $\bfq_m^J$ and can be found in Appendix \ref{appb}.

The mean field Hamiltonian in \eqref{modelH} has similarities to the Bistritzer-Macdonald model describing the non-interacting bilayer of graphene \cite{Bistritzer_PNAS2011} and possesses similar symmetries: $C_6$ rotations, moir\'e translations $T_{\rm M}$, and the time reversal. When $\chi_{\rm AA} \neq \chi_{\rm BB}$, the $C_2$ symmetry is broken, and the $K^{\rm M}$ and $K^{\rm M'}$ points located in the corners of the hexagonal moir\'e Brillouin zone become gapped. Another symmetry-breaking pattern corresponds to $\bfq^J_n \neq 0$. In the latter phase, if simultaneously $\delta \neq 0$, the moir\'e translation symmetry $T_{\rm M}$ is broken, and the unit cell in the real space triples as shown with white in \autoref{plot1}(b). 


The correlated twisted bilayer model has some distinctive features compared to the non-interacting counterpart, which are captured by the self-consistency equations determining $\chi_{\rm AA}$, $\chi_{\rm BB}$, and $\chi_{\rm AB}$ (eq.~\ref{sc1}) and (\ref{sc2}). The Fourier transform of the hopping term with momentum $\bfK$ equal to the distance to the $K$-point acts as an inter-layer hybridization parameter $t'$ in the non-interacting Bistritzer-Macdonald model. 
Similarly, in our model, the corresponding superexchange $J'$ is found as the Fourier transform of the function $J(|i-j|)$ at the momentum $\bfK$. Besides $t'$ and $J'$, the mean-field values of $\chi_{\rm AA(\rm BB)}$ depend on the Fourier transforms of $J(|i-j|)$ but at different momenta equal to either zero or to the reciprocal lattice vector
\begin{equation}
    J_0'  = \int \frac{d^2 \bfr}{\Omega}  J(\bfr)\, , ~~~
    \bar J'  = \int \frac{d^2 \bfr}{\Omega} J(\bfr) e^{-i \bfG \cdot \bfr}\, .
\end{equation}
As shown in \autoref{appb}, the magnitudes of the order parameters $\chi_{\rm AA, \, BB}$ are proportional to the ratios $J'/J_0'$ and $J'/\bar J'$. Therefore, the resulting phase diagram depends on the spread of the superexchange: the more it is localized in the real space, the larger the interlayer spinon condensates are, and the $C_2$-breaking phases are stabilized at the smaller values of $J'$.

 We consider two types of ansatz for the interlayer superexchange terms $\hat H_{\rm inter}^J$: the uniform with $\bfq^J_n = 0$, and the $\bfq^J_n = \bfd_n$. The latter choice triples the unit cell which leads to the hybridization of two neighboring Dirac cones in different layers when  $\chi_{AA} \neq \chi_{BB}$. This configuration is expected to be energetically favorable at large values of $J'$.


\section{Results and Discussion}
\label{sec3}

\begin{figure}[t!tbp]
    \centering
    \includegraphics[width=0.85\textwidth]{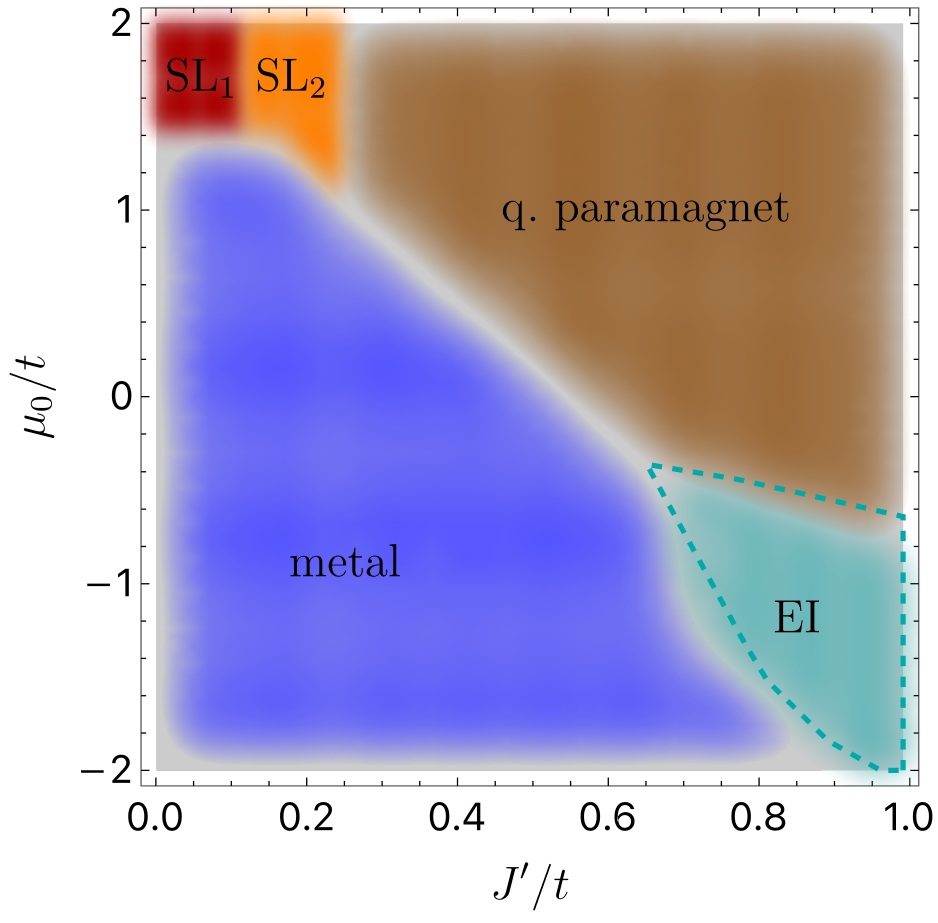}
    \caption{The phase diagram for the hexagonal Mott insulator bilayer with the twist angle $\theta = 10^\circ$. The parameters used are $t'=t''=t/5$, $J=0.5 t$, $J''=0.1 t$, $J_0' = J_0'' =  0.2 t$, $\bar J' = 0$. The blue dashed line indicates the change in the boundary of the EI phase when the twist angle is lowered to $7^\circ$.}
    \label{diag}
\end{figure}

By minimizing the Hamiltonian \eqref{modelH} with respect to the free parameters $\lambda$, $\delta$, $\chi$, $\chi_{\rm AA}$, $\chi_{\rm BB}$, $\chi_{\rm AB}$ and solving the resulting mean-field equations (see Appendix \ref{appb}) self-consistently, we obtain the phase diagram as presented in \autoref{diag} in terms of the interlayer superexchange $J'$ and the chemical potential $\mu_0$. The breakdown of the phases with associated order parameters and symmetry properties is shown in \autoref{phasetab}.

For small values of $J'$, the ground state corresponds to the uniform order parameter $\chi_{\rm AA} = \chi_{\rm BB}$, i.e. $\bfq^J = 0$. This metallic phase, highlighted in dark blue, exhibits a dispersion that is essentially a renormalized version of that found in twisted bilayer graphene. Yet there are key differences such as the non-linear dependence of the magic angle as a function of parameters as discussed below.

For higher values of the chemical potential, $\mu_0$, we obtain Mott insulating phases (i.e. no doping, $\delta=0$). Within the parton mean-field theory, these phases are either the spin liquids or the quantum paramagnets. The spin liquid phases are highlighted in red and orange in the phase diagram: the first phase is the plain intralayer $U(1)\times U(1)$ spin liquid with $\chi \neq 0$ and all other order parameters being zero. The latter phase highlighted with orange is a $U(1)$ spin liquid state with the gap at $E = 0$ arising due to the development of the non-trivial interlayer spinon order parameter $\chi_{\rm AA} = - \chi_{\rm BB}$ with $\bfq_n^J = \bfd_n$. Due to the gap opening, this phase features the chiral edge currents of spinons localized on the boundaries between the AB and BA stacking arrangements \cite{network}. For larger values of interlayer Heisenberg exchange, $J'$, the Dirac kinetic term provided by the intralayer spinon order parameter vanishes with the corresponding phase becoming a topologically trivial flat band. This gapped phase which we call the quantum paramagnet and show with brown color is characterized by the formation of interlayer singlets.

\begin{figure}[t] 
    \centering
        \centering
        \includegraphics[width=\textwidth]{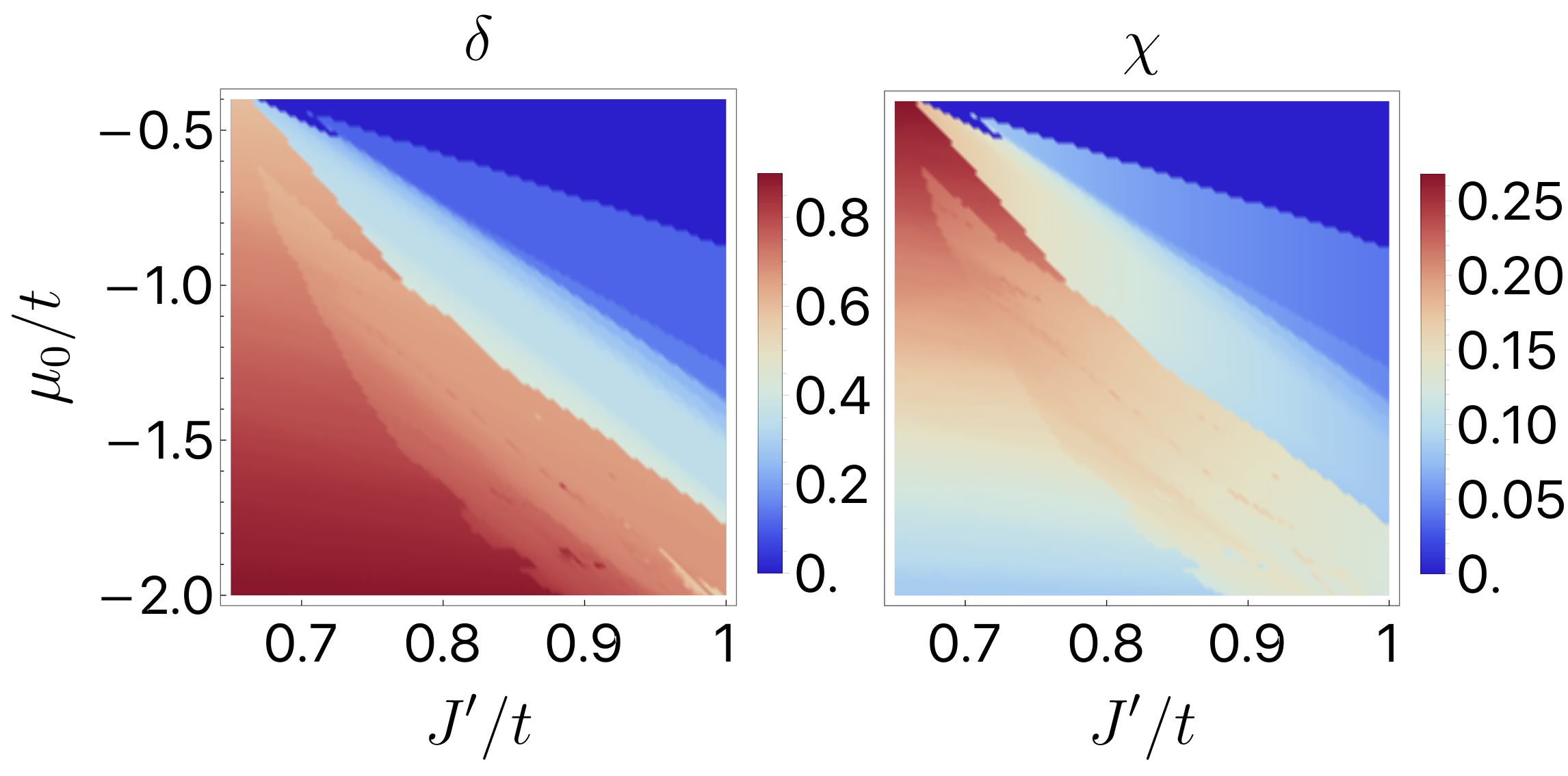}
        \label{plot3}
    \caption{Mean-field parameters $\delta$ and $\chi$ in the vicinity of the excitonic insulator phase in the bottom right corner of the phase diagram in \autoref{diag}. Both quantities experience abrupt changes as $J'$ and $\mu_0$ are varied when several minibands cross the Fermi level.}
    \label{ei}
\end{figure}

The last phase, shown with cyan, is insulating with $\delta > 0$ and is particularly distinctive compared to the non-interacting twisted models. The Fermi level in this phase lies between the minibands with the mean-field parameters $\delta$ and $\chi$ varying across the phase in a step-like fashion, as shown in \autoref{ei}. Thus, the latter phase encompasses multiple phases separated by the metal-insulator transitions. As one can infer from the \autoref{phasetab}, in this phase both the interlayer and intralayer spinon condensates are present. Furthermore, the combination of $\delta \neq 0$ and the momentum dependence of the order parameter breaks spontaneously the moir\'e translational symmetry, such that the area of the unit cell in the real space triples, as shown in \autoref{plot1}($b$). Therefore, this state is an exotic valence bond solid characterized by the formation of the inter-layer singlets.

As indicated in \autoref{diag} with the blue dashed line, the excitonic insulator phase expands towards $\mu_0 = 0$ as the twist angle is lowered, due to the increase in the number of minibands. Therefore, at small twist angles, this phase is realized at a lower doping level. However, crucially, we find that the condition $J' \gg J_0'$ needs to be satisfied in order for the EI phase to exist, i.e. the superexchange at zero momentum should be much smaller than at the momentum $\bfK$. As a result, some destructive interference effects with respect to the nearest-neighbor superexchange are likely required to observe this phase. Lastly, we note that the mean-field parameters that distinguish different phases are the interlayer spinon susceptibilities $\chi_{\rm AA}$ and $\chi_{\rm BB}$. The parameter $\chi_{\rm AB}$ is found to be non-zero only in the metal phase.

\begin{table}[t!tbp]
\centering

\begin{tabular}{ |p{4cm}|p{0.3 cm}|p{0.3cm}|p{0.6cm}|p{0.6cm}|p{0.4cm}|p{0.4cm}|p{0.4cm}|}
 \hline
 \multicolumn{8}{|c|}{Summary of the phases} \\
 \hline
 Phase & $\d$ & $\chi$ & $\chi_{\rm AA}$ &$\chi_{\rm BB}$ & $\bfq_n^J$ & $C_2$ & $T_{\rm M}$ \\
\hline
 \redbox[2pt]{} $U(1)\times U(1)$ Spin Liquid & 0 & $+$ & 0 &  0 & $\bfd_n$ & \textcolor{darkgreen}{\CheckmarkBold} & \textcolor{darkgreen}{\CheckmarkBold}  \\
\hline
 \orangebox[2pt]{} $U(1)$ Gapped Spin Liquid  & 0 & $+$ & $+$ & $-$ & $\bfd_n$ & \textcolor{darkred}{\XSolidBrush} & \textcolor{darkgreen}{\CheckmarkBold}   \\
\hline
\brownbox[2pt]{} Quantum Paramagnet    & 0 & 0 & $+$ & $-$ & $\bfd_n$ & \textcolor{darkred}{\XSolidBrush} & \textcolor{darkgreen}{\CheckmarkBold}  \\
 \hline
 \bluebox[2pt]{} Metal & $+$ & $+$ & $+$ & $+$ & 0 & \textcolor{darkgreen}{\CheckmarkBold} & \textcolor{darkgreen}{\CheckmarkBold}  \\
 \hline
 \cyanbox[2pt]{} Excitonic Insulator (EI) & $+$ & $+$ & $+$ & $-$ & $\bfd_n$ & \textcolor{darkred}{\XSolidBrush} & \textcolor{darkred}{\XSolidBrush}  \\
\hline
\end{tabular}
\caption{The first column labels the phase together with its color code in the diagram \autoref{diag}. The next four columns represent the values of the parameters distinguishing the phases in the diagram. The sign ``$+$'' means that the order parameter is positive and ``$-$'' stands for negative. The sixth column indicates whether the interlayer spinon order parameters $\chi_{\rm AA(\rm BB)}$ are uniform ($\bfq_n^J = 0$) or develop spatial modulation ($\bfq_n^J = \bfd_n$). The last two columns show whether the $C_2$ symmetry protecting the Dirac cones and the moir\'e translations $T_{\rm M}$ are spontaneously broken (indicated by the cross mark) or preserved (shown with the check mark).}
\label{phasetab}
\end{table}

Another distinctive aspect of the considered model is the superexchange dependence of the magic angles found in the correlated metallic phase. As discussed above, the Hamiltonian in this phase is in many ways analogous to the Bistritzer-Macdonald model with certain parameters determined self-consistently. The magic angles then occur due to the interplay between the interactions and the dispersion in turn both dependent on mean-field parameters such as $\chi$ and $\delta$.

\begin{figure}[h!tbp]
         \centering       \includegraphics[width=1\textwidth]{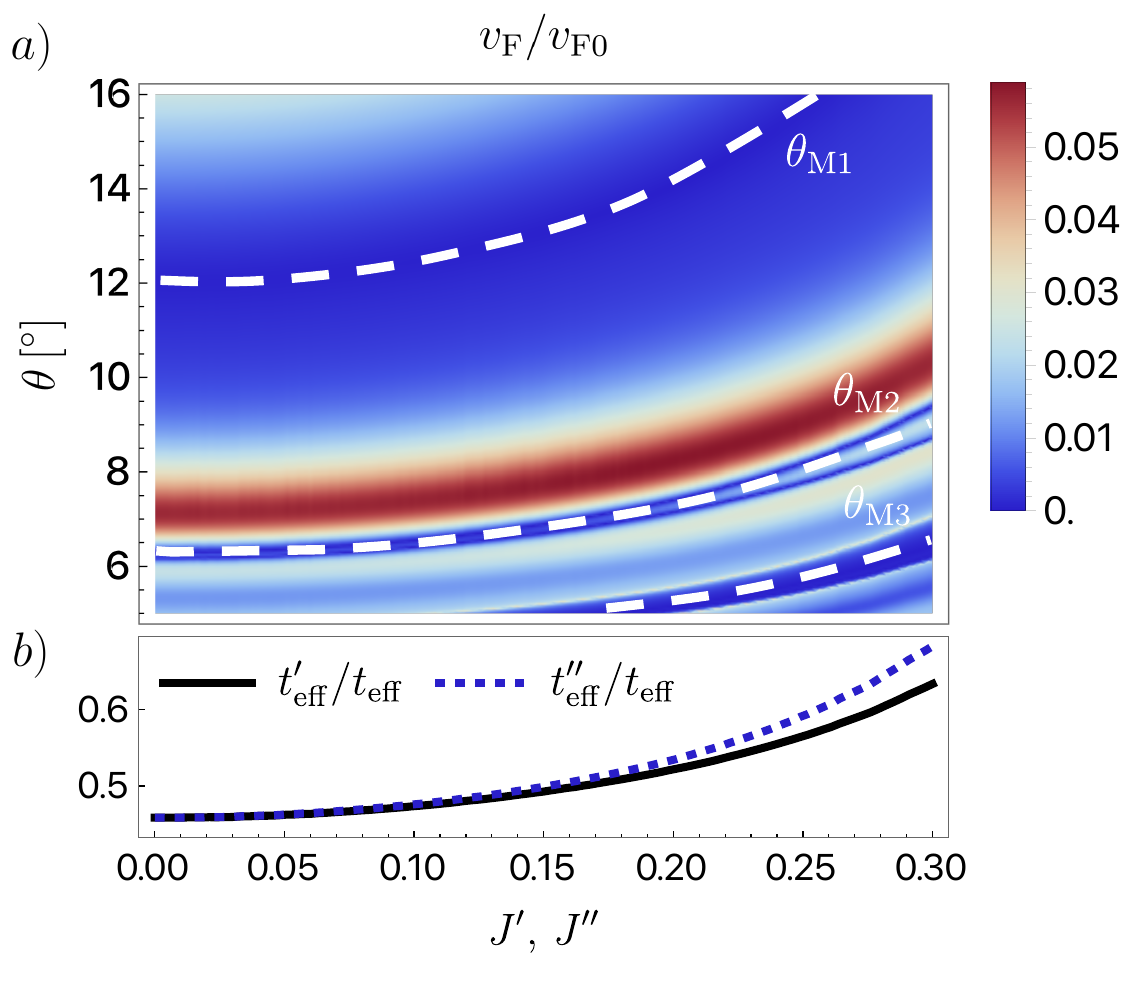}
        \caption{Renormalization of the three largest magic angles of spinons ($a$) and the effective tunneling terms ($b$) as a function of parameters $J'$ and $J''$ as they are varied simultaneously. Due to the net increase in the interlayer tunneling with respect to the intralayer shown on the bottom panel, the magic angles increase as indicated on the Fermi velocity density plot on the top panel. The parameters used are: $t'=t''=t/2$, $J=0.2  t$, $J_0' = J_0'' = t/4$, $\bar J' = 0$, and $v_{\rm F0} = 3a t/2$ denotes the bare Fermi velocity}
         \label{mangle}
\end{figure}

In \autoref{mangle} $a)$, we showcase the effect of the interlayer superexchanges $J'$, $J''$ on the three largest magic angles. The main influence on the dispersion of the energy levels at neutrality can be captured via the renormalization of the effective intralayer and two interlayer tunneling terms 
\begin{equation}
\begin{aligned}
t_{\rm eff} & = \delta t + \frac{\chi J}{4} \, , \\ t'_{\rm eff} & = \delta t' + \frac{\chi_{\rm AA( BB)} J'}{4} \, , \\ t''_{\rm eff} & = \delta t'' + \frac{\chi_{\rm AB} J''}{4} \, .
\end{aligned}
\label{teff}
\end{equation}
From \autoref{mangle} $b)$, we infer that when the parameters $J'$, $J''$ are increased, the ratios $t'_{\rm eff}/t_{\rm eff}$ and $t''_{\rm eff}/t_{\rm eff}$ also grow. This is to be expected from the definitions \eqref{teff}, however, the actual increase is faster than linear due to the mean-field renormalization of $\chi_{\rm AA}$, $\chi_{\rm BB}$ and $\chi_{\rm AB}$. Since the effective interlayer tunneling is increased, all magic angles increase in magnitude.

\section{Conclusion}
\label{sec4}

We examined the twisted hexagonal correlated insulators via the slave boson mean field theory. The associated phase diagram, which now results from the effect of twist on both the band structure and the form of interactions, is shown to host the exotic translational symmetry-breaking excitonic insulator phase. This phase arises due to the existence of the mini-gaps controlled by the interaction, and hence sensitive to the twist angle. Despite the similarity of our model Hamiltonian at the mean-field level with the BM model, the interactions play an essential role in the stabilization of phases that are absent in the non-interacting BM model. In particular, the flat bands corresponding to magic angles are determined by the electron-electron interactions as much as the tunneling terms between the layers. Furthermore, due to the correlated nature of underlying monolayers, changes in doping levels have a significant impact on the parameters of the model. The high tunability of the lower-dimensional materials with respect to doping raises hopes that the phenomenology described in this work can be observed in future experiments.

\begin{acknowledgments}
We thank Arun Paramekanti and Liuyan Zhao for the fruitful discussions. PG acknowledges support from National Science Foundation Award NO. DMR-2130544. This work was supported by the U.S. Department of Energy, Office of Science, Office of Basic Energy Sciences, Material Sciences and Engineering Division under Award Number DE-SC0025247. The work of IK is supported by Julian Schwinger Foundation.
\end{acknowledgments}

\appendix


\section{Mean field theory for AA-stacking pattern}

\label{appa}

In this appendix, we derive the mean-field formalism for the AA-stacked bilayer. We assume the interlayer hopping to be local and the AB hopping absent. In the limit of $U\rightarrow \infty$ and close to half-filling, the $t-J$ model is a valid approximation
\begin{eqnarray}
   \hat H_{t-J} =& -& \sum_{\braket{ij}} t_{ij} \hat P \hat c^\dagger_{i \s} \hat c_{j \s} \hat P  -\mu_0 \sum_i \hat c^{\dagger}_{i \s} \hat c_{i \s}\, \nonumber\\
  & +& \sum_{\braket{ij}} J_{ij} (\hat \bfS_i \cdot \hat \bfS_j - \frac{1}{4} \hat n_i \hat n_j)
\label{tjapp} 
\end{eqnarray}
where $t_{ij}$ denotes the hopping between the sites $i$ and $j$ (both of the indices may take values in the first and the second layers), $J_{ij}$ is the superexchange parameter, and $\hat P$ is the projector into the space of singly-occupied sites. The operators $\hat c_{i\sigma}$ and $\hat c_{i\sigma}^\dagger$ are the annihilation and creation operators of fermions with position $i$ and spin $\sigma$. In order to accommodate the constraint of not having doubly occupied sites, we use the parton decomposition, $\hat c_{i\sigma} = \hat b^\dagger_i \hat f_{i\sigma}\, $.
The constraint takes the convenient for numerical calculations holonomic form $\sum_\sigma \hat f^\dagger_{i\sigma} \hat f_{i\sigma}+\hat b^\dagger_i \hat b_i = 1 \, $.
In terms of the parton fields, the $t-J$ model Hamiltonian \eqref{tjapp} with the added constraint reads \cite{lee2004}
\begin{eqnarray}
  \hat H_0 = &&- t \sum_{\braket{ij}} (\hat f^\dagger_{i \s} \hat f_{j \s} \hat b_i \hat b^{\dagger}_j + {\rm{c. c.}}) - \mu_0 \sum_i \hat f^{\dagger}_{i \s} \hat f_{i \s} \nonumber\\
  &&+ J \sum_{\braket{ij}}(\hat \bfS_i \cdot \hat \bfS_j - \frac{1}{4} \hat n_i \hat n_j) \nonumber\\
  &&+ \l \sum_i \( \hat f^{\dagger}_{i \s} \hat f_{i \s} + \hat b^\dagger_i \hat b_i - 1 \)\, .
\label{tjmonoapp}   
\end{eqnarray}
We assume constant mean occupation of holes $\braket{\hat b_i^\dagger \hat b_i} = \d$ and decouple the spinon terms in the exchange channel introducing the real-valued Hubbard-Stratonovich field $\chi_{i j} = \braket{\hat f^\dagger_{i \s} \hat f_{j \s}} \d_{\braket{i j}}$ that modifies the nearest-neighbor hoppings. The mean-field expressions for the density-density and spin-spin terms then read
\begin{align}
\hat n_i \hat n_j & =  (1-\hat b^\dagger_i \hat b_i)(1-\hat b^\dagger_j \hat b_j) \sim (1-\d)^2 \, , \nonumber \\
\hat \bfS_i \cdot \hat \bfS_j & =  2 \hat f^\dagger_{i \s} \hat f_{i \b} \hat f^\dagger_{j \b} \hat f_{j \s} - (\hat f^\dagger_{i \s} \hat f_{i \s})(\hat f^\dagger_{j \b} f_{j \b}) \label{nnssapp} \\
~ & \sim \frac{1}{4}\chi_{ij}^2-\frac{1}{4}(\chi_{i j} f^\dagger_{\s i} f_{\s j} + {\rm c. c.}) + \frac{1}{4}(1-\d)\, . \nonumber
\end{align}
The proof of the last identity is contained in \autoref{suba}. By utilizing the identities \eqref{nnssapp} in \eqref{tjmonoapp}, and taking the Fourier transform, we obtain the Hamiltonian for a single monolayer
\begin{widetext}
\begin{equation}
\hat H_{0}= - (\d t + \frac{J \chi}{4}) \sum_{\bfk\s} (g(\bfk) \hat f^\dagger_{\bfk \s \rm A} \hat f_{\bfk \s \rm B} + {\rm c. c.}) + (\l - \mu_0) \sum_{\bfk \s \alpha} \hat f^\dagger_{ \bfk \s \a} \hat f_{\bfk \s \alpha} + \frac{3 J N}{4} \chi^2 + \frac{N}{4} (1-\d) (3 \d J - 8 \l) \, ,
\label{h0app}
\end{equation}
\end{widetext}
where $N$ is the number of unit cells, and the function $g(\bfk) = \sum_{j} e^{- i \bfk \cdot \bfg_j} $ is the hopping term well-known from the theory of graphene monolayer: in our notations $\bfg_1 = (0, a)$ with the remaining $\bfg_{j}$ obtained by $2 \pi/3$ rotations. In order to describe a strongly interacting bilayer, we take the Hamiltonian \eqref{h0app} twice---for the layers $1$ and $2$---and supplement the sum by the interlayer hopping and interlayer superexchange 
\begin{equation}
    \hat H_{\rm AA} = \hat H_{0}^{(1)} + \hat H_{0}^{(2)} + \hat H_{t'} + \hat H_{J'} \, ,
\end{equation}
where in the basis 
\beq
\hat f_\bfk = (\hat f^{(1)}_{\bfk A} ~~ \hat f^{(1)}_{\bfk B} ~~ \hat f^{(2)}_{\bfk A} ~~ \hat f^{(2)}_{\bfk B}) \, ,
\eeq
we write\\\\\\
\beq
H_{t'} = - \begin{pmatrix}
0 & 0 & 3\d  t' & 0 \\
0 & 0 & 0 & 3\d  t' \\
3\d  t' & 0 & 0 & 0 \\
0 & 3\d  t' & 0 & 0
\end{pmatrix} \, ,
\eeq
with the hopping $t'$ tripled to make the comparison with the non-zero twist angle case more convenient. We use $3 J'$ for the superexchange alike. The interlayer superexchange contains spin-spin and density-density interlayer terms
\begin{eqnarray}
\hat H_{J'} = 3 J' \sum_i \(  \hat\bfS^{(1)}_{i \rm A} \cdot \hat\bfS^{(2)}_{i \rm A} +  \hat\bfS^{(1)}_{i \rm B} \cdot \hat\bfS^{(2)}_{i \rm B} \) \nonumber \\
-3 J'\sum_i \( \frac{1}{4} \hat n^{(1)}_{i \rm A} \hat n^{(2)}_{i \rm A}  - \frac{1}{4} \hat n^{(1)}_{i \rm B} \hat n^{(2)}_{i \rm B}  \)
\end{eqnarray}

Making a further assumption that the hole occupation is the same in each layer and sublattice, all above terms can be treated using the expressions analogous to \eqref{nnssapp}. The resulting Hamiltonian for the AA-stacking reads
\begin{widetext}
\beq
\hat H_{\rm untw} = - \sum_{\bfk\sigma} \hat f_{\bfk\sigma}^\dagger \begin{pmatrix}
\mu_0 - \l & (\d t + \frac{\chi J}{4} )g(\bfk) & 3\d t' + \frac{3\chi_{\rm AA} J'}{4} & 0 \\
(\d t + \frac{\chi J}{4})g^*(\bfk) & \mu_0 - \l & 0 & 3\d t' + \frac{3\chi_{\rm BB} J'}{4} \\ 
 3\d t' + \frac{3\chi_{\rm AA} J'}{4} & 0 & \mu_0 - \l & (\d t + \frac{\chi J}{4}) g(\bfk) \\
 0 & 3\d t' + \frac{3\chi_{\rm BB} J'}{4} &  (\d t + \frac{\chi J}{4}) g^*(\bfk) & \mu_0 - \l
\end{pmatrix} \hat f_{\bfk\sigma} + H_{\rm const} \, ,
\label{haaapp}
\eeq
\end{widetext}
where the constant term is $H_{{\rm const}} = \frac{3}{2} N J \chi^2 + \frac{3 N J'}{4}(\chi_{\rm AA}^2+\chi_{\rm BB}^2) + \frac{N}{2}(1-\d)(3 \d J + 3 \d J' - 8 \l ) \,$. The parameters of the Hamiltonian are $\mu_0$, the intra- and interlayer hoppings $t$, $t'$, and intra- and interlayer superexchanges $J$, $J'$. The mean-field parameters $\chi$, $\chi_A$, $\chi_B$, $\lambda$ and $\delta$ are self-consistently determined by minimizing the free energy
\begin{widetext}
\beq
\bal
\chi & = \frac{1}{6} \int_{\rm BZ} \frac{d^2 \bfk}{K} \sum_{n=1}^4 \braket{n\bfk|\hat H_0|n\bfk} \theta \(E_F - E_{n}(\bfk)\) \equiv \frac{I_0}{6} \, , \\
\chi_{{\rm AA(BB)}} & = \int_{\rm BZ} \frac{d^2 \bfk}{K} \sum_{n=1}^4 \braket{n\bfk|\hat H_{{\rm AA(BB)}}|n\bfk} \theta \(E_F- E_{n}(\bfk)\) \equiv I_{AA(BB)} \, , \\
\d & = 1 - \frac{1}{2} \int_{\rm BZ} \frac{d^2 \bfk}{K} \sum_{n=1}^4 \theta \(E_F- E_{n}(\bfk)\) \, , \\
\l & = \frac{1}{2} t I_0 + \frac{3 t'}{2}(I_{\rm AA} + I_{BB})  - \frac{3}{8}(1-2 \d)( J + J') \, , 
\eal
\label{mfequntwapp}
\eeq
\end{widetext}
where $\hat H_0 = \sum_{\bfk l}g(\bfk) \hat f_{\bfk\rm A}^{(l)\dagger} \hat f_{\bfk \rm B}^{(l)} + {\rm h.c.}$, $\hat H_{\rm AA(BB)} = \sum_{\bfk} \hat f_{\bfk \rm A(B)}^{(1)\dagger} \hat f_{\bfk \rm A(B)}^{(2)} + {\rm h.c.}$, $E_n(\bfk)$ and $\ket{n \bfk}$ are the eigenvalues and eigenvectors of $\hat H_{\rm untw}$, $K$ is the area of the hexagonal Brillouin zone, and $\theta$ denotes a step function.

\subsection{Mean-field treatment of $\hat \bfS_i \cdot \hat \bfS_j$ term}

\label{suba}

We start by recalling the identity for Pauli matrices $\sigma_{\alpha \beta} \sigma_{\gamma \delta}=2 \delta_{\alpha \delta} \delta_{\beta \gamma}-\delta_{\alpha \beta} \delta_{\gamma \delta}$. With this, the spin exchange can be cast as
\begin{equation}
4 \left( \hat \bfS_i \cdot \hat \bfS_j \right)=2 \hat f_{\sigma_i}^{\dagger} \hat f_{\beta i} \hat f_{\beta j}^{\dagger} \hat f_{\sigma j}- \hat f_{\sigma_i}^{\dagger} \hat f_{\sigma_i} \hat f_{\beta j}^{\dagger} \hat f_{\beta j}
\end{equation}
The first term on the right-hand side we write as
\begin{equation}
2 \hat f_{\sigma i}^{\dagger} \hat f_{\sigma j} \hat f_{\beta i} \hat f_{\beta j}^{\dagger}+2 \hat n_i  \\
= -2 \hat f_{\sigma i}^{\dagger} \hat f_{\sigma j} \hat f_{\beta j}^{\dagger} \hat f_{\beta i}+2 \hat n_i \, .
\end{equation}
The remaining term is expanded as
\begin{equation}
\begin{gathered}
    \hat f_{\uparrow i}^{\dagger} \hat f_{\uparrow i} \hat f_{\uparrow j}^{\dagger} \hat f_{\uparrow j}+\hat f_{\downarrow i}^{\dagger} \hat f_{\downarrow i} \hat f_{\downarrow j}^{\dagger} \hat f_{\downarrow j}\\
    -\hat f_{\uparrow i}^{\dagger} \hat f_{\uparrow j} \hat f_{\downarrow j}^{\dagger} \hat f_{\downarrow i}-\hat f_{\downarrow i}^{\dagger} \hat f_{\downarrow j} \hat f_{\uparrow j}^{\dagger} \hat f_{\uparrow i}\\
    +\hat f_{\uparrow i}^{\dagger} \hat f_{\uparrow j} \hat f_{\downarrow j}^{\dagger} \hat f_{\downarrow i}+\hat f_{\downarrow i}^{\dagger} \hat f_{\downarrow j} \hat f_{\uparrow j}^{\dagger} \hat f_{\uparrow i}\\
+\hat f_{\uparrow i}^{\dagger} \hat f_{\uparrow i} \hat f_{\downarrow j}^{\dagger} \hat f_{\downarrow j}+\hat f_{\downarrow i}^{\dagger} \hat f_{\downarrow i} \hat f_{\uparrow j}^{\dagger} \hat f_{\uparrow j}\, ,
\end{gathered}
\end{equation}
where in the second line we added and in the third line removed the same two terms. Using the anti-commutation relations, $\left\{\hat f^\dagger_{i\alpha}, \hat f_{j\beta}\right\} = \delta_{ij}\delta_{\alpha\beta}$, we move all the operators with a dagger in the last two lines to the left, which gives overall
\begin{eqnarray}
    4 \left( \hat \bfS_i \cdot \hat \bfS_j \right)&= &\hat n_{i \uparrow}+\hat n_{i \downarrow}-\hat f_{\sigma_i}^{\dagger} \hat f_{\sigma j} \hat f_{\beta j}^{\dagger} \hat f_{\beta i}- \hat f_{\uparrow i}^\dagger \hat f_{\downarrow j}^\dagger \hat f_{\uparrow j} \hat f_{\downarrow i} \nonumber \\
    &&-\hat f_{\downarrow i}^\dagger \hat f_{\uparrow j}^\dagger \hat f_{\downarrow j} \hat f_{\uparrow i} + \hat f_{\uparrow i}^\dagger \hat f_{\downarrow j}^\dagger \hat f_{\downarrow j} \hat f_{\uparrow i} \nonumber \\
    &&+ \hat f_{\downarrow i}^\dagger \hat f_{\uparrow j}^\dagger \hat f_{\uparrow j} \hat f_{\downarrow i} \, .
\end{eqnarray}
The last four terms combine into the pairing term $\left(\hat f_{\uparrow i}^{\dagger} \hat f_{\downarrow j}^{\dagger}-\hat f_{\downarrow i}^{\dagger} \hat f_{\uparrow j}^{\dagger}\right)\left(\hat f_{\downarrow j} \hat f_{\uparrow i}-\hat f_{\uparrow j} \hat f_{\downarrow i}\right)$ that we ignore in this work. Thus, we obtain the final expression
$\hat \bfS_i \cdot \hat \bfS_j \simeq-\frac{1}{4} \hat f_{\alpha i}^{\dagger} \hat f_{\alpha j} \hat f_{\beta j}^{\dagger} \hat f_{\beta i}+\frac{1}{4} \hat n_i \sim 
-\chi_{i j} \hat f_{\beta j}^{\dagger} \hat f_{\beta i}-\chi_{i j} \hat f_{\beta i}^{\dagger} \hat f_{\beta j}+\chi_{i j}^2 + \frac{1}{4} \hat n_i \, .$

\section{Derivation of the twisted bilayer mean-field Hamiltonian}

\label{appb}

\begin{figure}[h!tbp]
    \centering
\includegraphics[width=1\linewidth]{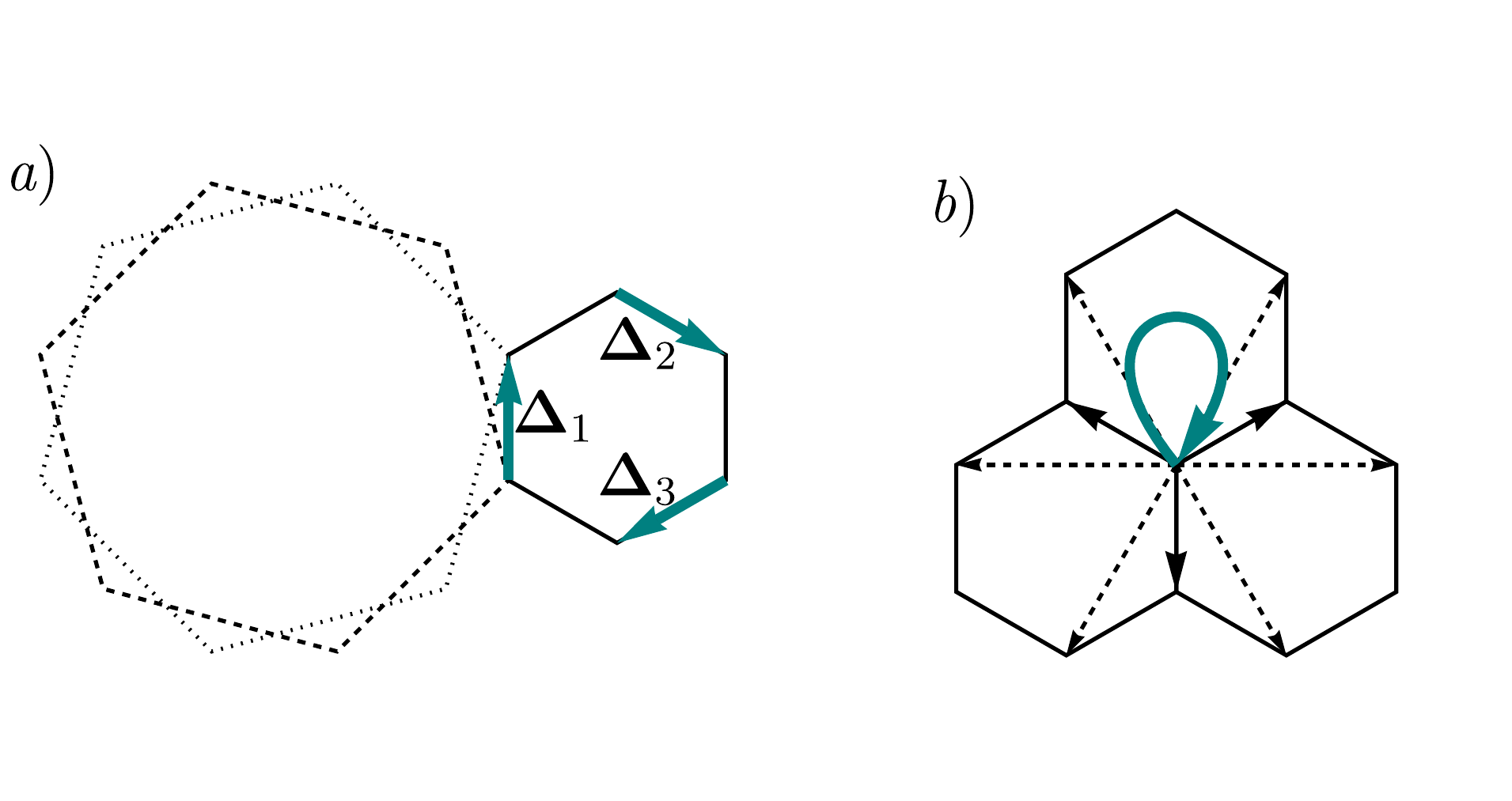}
    \caption{$a)$: the small hexagon shows the moir\'e Brillouin zone; vectors $\bfd_n$ connecting vertices on the moir\'e reciprocal lattice are indicated with blue arrows. $b)$: the allowed hybridization momenta in the twisted bilayer model with $\bfq_n^J \neq 0$. Each solid hexagon indicates the moir\'e Brillouin zone. Note that as indicated by the blue arrow, every Dirac cone can hybridize with the Dirac cone from the other layer at the same point in the momentum space.}
    \label{hexplots}
\end{figure}

In this appendix, we derive the effective continuous model of the twisted strongly correlated bilayer. In the limit of zero twist angle, we expect it to reproduce the continuum limit of the Hamiltonian \eqref{haaapp}.

We use the indices $a, b=1,2$ to denote the layer index, and $\alpha, \beta = \rm A,B$ to label sublattices. The layers $1$ and $2$ are rotated by the angles $\theta/2$ and $-\theta/2$ with respect to an imaginary reference layer. The reciprocal lattice vectors in the reference layer are given by $\bfG_{1,2} =  2 \pi/(\sqrt{3}a)(-1, \mp 1/\sqrt{3})$. The moir\'e reciprocal lattice vectors in the small twist angle limit are $\bfG_{1,2}^{ \rm M} = 2 \pi/(\sqrt{3}a)  (\pm 1/\sqrt{3}, -1) \theta$ (we assume throughout that $\theta$ is small). The vectors $\bfd_n$ connect nearest neighbors in the moir\'e lattice, with $\bfd_1 =(0, 4 \pi \theta/(3 \sqrt{3}a))$, and $\bfd_{2,3}$ obtained from it by $2 \pi/3$ and $4 \pi/3$ clockwise rotations (see the left panel in \autoref{hexplots}). The moir\'e reciprocal lattice vectors $\bfG^{\rm M}$ and $\bfd_n$ are related as 
\begin{equation}
    \bfG_1^{\rm M} = \bfd_2 - \bfd_1\, , \qquad \bfG_2^{\rm M} = \bfd_3 - \bfd_1 \, .
\end{equation}

The derivation below closely follows one of the original continuous Bistritzer-Macdonald model \cite{bernevig}. The starting assumption is that one can write the $t-J$ model \eqref{tjapp} in the Wannier basis of the original monolayers. The form of the interlayer terms in the resulting parton mean field theory will essentially depend on the properties of the interlayer superexchange parameter $J(|i-j|)$. It is clear that if the layers are sufficiently far away from each other, the layer corrugation is smooth, and the spatial variation of $J$ is also expected to be smooth. With this assumption, a Bistritzer Macdonald (BM)-type model can be derived, as we demonstrate below.

After switching from fermion to spinon fields, and introducing the mean-field parameter $\chi_{ij}$ as explained in \autoref{appa}, one obtains a mean-field Hamiltonian analogous to \eqref{haaapp}. The essential difference is that in the non-zero twist angle case, the superexchange is no longer constant in space, which leads to the appearance of the terms
\begin{align}
   \hat U &= -\frac{1}{4}\sum_{ij} J(|i-j|) \chi_{ij} \hat f^{\dagger(1)}_i \hat f_j^{(2)} \, , \nonumber \\ 
   \hat U^* &= -\frac{1}{4}\sum_{ij} J(|i-j|) \chi^*_{ij} \hat f^{\dagger(1)}_j \hat f_i^{(2)}\, , \nonumber \\ 
   X &= \frac{1}{4}\sum_{ij} J(|i-j|) \chi^*_{ij} \chi_{ij} \, , \nonumber \\
   { \mathcal{N}} &= - \frac{1}{4} \sum_{ij} J(|i-j|) \hat{n}_i \hat{n}_j \, ,
   \label{termsq}
\end{align}
in place of the superexchange terms in \eqref{haaapp}. We further assume that the mean field $\chi_{ij}$ can condense at non-zero momenta $\bfq^J_n$
\beq
\chi_{ij}= \chi_{\a \b} \sum_{n=1}^{3} e^{i \bfd_n \cdot \bfj}\, .
\label{chiansapp}
\eeq
The reason to expect $\bfq^J_n \neq 0$ is that when the interlayer superexchange is sufficiently high, the gap opening is energetically favorable. In order for the gap to open, the Dirac cones belonging to different layers need to be able to hybridize at zero momentum. This is achieved precisely by setting $\bfq^J_n = \bfd_n$, which will be evident from the form of the Hamiltonian obtained below.

We begin our BM-type model derivation by evaluating the expectation value of the first ($\hat U$) term in \eqref{termsq} between the monolayer Bloch states
\beq
\ket{f^{(a)}_{\bfp \a}} = \frac{1}{\sqrt{N}} \sum_{\bfR^{(a)}} e^{i \bfp \cdot (\bfR^{(a)} + \bft^{(a)}_{\a})} \ket{\bfR^{(a)} + \bft_{\a}^{(a)}}\, ,
\eeq
where $\bfR^{(a)}$ labels the positions of atoms on A sublattice in layer $(a)$, and $\bft_\a^{(a)}$ is the basis vector in the same layer: $\bft_{\rm A} = (0,0)$, $\bft_{\rm B} = (a,0)$ before twisting. The requisite interlayer matrix element can be written as
\begin{widetext}
\beq
\begin{aligned}
U_{\a\b}^{(12)}(\bfp,\bfp') \equiv \braket{f^{(1)}_{\bfp \a}|\hat U|f^{(2)}_{\bfp' \b}} = - \f{\chi_{\a \b}}{4N} \sum_{\bfR^{(1)} \bfR^{(2)} n} e^{-i \bfp \cdot (\bfR^{(1)}+\bft_\a^{(1)})} e^{i \bfp' \cdot (\bfR^{(2)}+\bft_\b^{(2)})} & e^{i \bfq^J_n \cdot (\bfR^{(2)}+ \bft^{(2)}_\b)} \\ \times & J(\bfR^{(1)}+\bft^{(1)}_\a-\bfR^{(2)} - \bft^{(2)}_\b)\, .
\end{aligned}
\label{hamst1}
\eeq
We further expand $J$ in a Fourier integral
\beq
J(\bfR^{(1)}+\bft^{(1)}_\a-\bfR^{(2)} - \bft^{(2)}_\b) = \f{1}{N \Omega} \sum_\bfq \sum_\bfG J_{\bfq+\bfG}^{(12)} e^{i (\bfq+\bfG) \cdot (\bfR^{(1)}+\bft_\a^{(1)}-\bfR^{(2)}-\bft_\b^{(2)})}\,,
\eeq
\end{widetext}
where the $\bfG$-summation runs over the reciprocal lattice vectors of the reference layer and $\bfq$-summation is over the hexagonal Brillouin zone. The letter $\Omega$ denotes the size of the unit cell in the real space, and $N$ is the number of electrons. Plugging this expression back into the matrix element \eqref{hamst1} produces
\begin{widetext}
\beq
\bal
U^{(12)}_{\a \b} (\bfp, \bfp') = - \frac{\chi_{\a \b}}{4 N^2 \Omega} & \sum_{\bfR^{(1)} \bfR^{(2)}} \sum_{\bfq \bfG n} e^{i (\bfq+\bfG-\bfp) \cdot \bfR^{(1)} } e^{- i (\bfq+\bfG-\bfp'-\bfq^J_n) \cdot  \bfR^{(2)}} J^{(12)}_{\bfq+\bfG} \\ \times & e^{i (\bfq^J_n \cdot \bft^{(2)}_\b+ \bfp' \cdot \bft_\b^{(2)}  - \bfp \cdot \bft^{(1)}_\a   + \bfq \cdot (\bft_\a^{(1)}-\bft_\b^{(2)})  +\bfG \cdot (\bft_\a^{(1)}-\bft_\b^{(2)}) )}\,.
\label{hamst2}
\eal
\eeq
We further make use of the identities
\beq
\bal
\f{1}{N} \sum_{\bfR^{(2)}} e^{i  (\bfq+\bfG-\bfp'-\bfq^J_n) \cdot \bfR^{(2)}} & = \sum_{\bfG^{(2)}} \d_{\bfq+\bfG,\bfp'+\bfq^J_n+\bfG^{(2)}} = \sum_{\bfG'} \d_{\bfp'-\bfq-\bfG+R_-\bfG',-\bfq^J_n} \, , \\
\f{1}{N} \sum_{\bfR^{(1)}} e^{i (\bfq + \bfG - \bfp) \cdot \bfR^{(1)}} & = \sum_{\bfG ''} \d_{\bfp - \bfq-\bfG + R_+ \bfG'',0} \, ,
\eal
\eeq
where $\bfG'$ and $\bfG''$ lie within the reference layer, $R_+$ is a rotation matrix with parameter $\theta/2$, and $R_- = R_+^T$. Integration over $\bfq$ present in \eqref{hamst2} combines these two delta-functions into the one with the argument
\beq
\bfp-\bfp'+R_+ \bfG''-R_- \bfG'- \bfq^J_n\, 
\eeq
with a condition $\bfq = \bfp' - \bfG+R_- \bfG' + \bfq^J_n$. Substituting these in \eqref{hamst2} gives
\beq
U^{(12)}_{\a \b} (\bfp, \bfp') =- \f{\chi_{\a \b}}{4\Omega} \sum_{\bfG \bfG' \bfG'' n} J^{(12)}_{\bfp' +R_- \bfG' + \bfq^J_n} \d_{\bfp-\bfp'+R_+ \bfG''-R_- \bfG'-\bfq^J_n}e^{i \k} \,,
\eeq
where the phase is
\beq
\bal
\k & = \bfp' \cdot \bft^{(2)}_\b   -  \bfp \cdot \bft^{(1)}_\a   + (\bft^{(1)}_\a - \bft^{(2)}_\b) \cdot (\bfp'+  R_- \bfG') + \bfq^J_n \cdot \bft^{(1)}_\a  \\
& = - \bfp \cdot \bft^{(1)}_\a   + \bfp' \cdot \bft^{(1)}_\a  +  (\bfp - \bfp' + R_+ \bfG''-\bfq^J_n) \cdot \bft^{(1)}_\a - R_- \bfG' \cdot \bft^{(2)}_\b  + \bfq^J_n \cdot \bft^{(1)}_\a   \\
& = R_+ \bfG'' \cdot  \bft^{(1)}_\a  - R_- \bfG' \cdot \bft^{(2)}_\b  \, .
\eal
\eeq
The matrix element then takes the form
\beq
U^{(12)}_{\a \b} (\bfp, \bfp') = - \f{\chi_{\a \b}}{4\Omega} \sum_{\bfG' \bfG'' n} J^{(12)}_{\bfp' +R_- \bfG' + \bfq^J_n} \d_{\bfp-\bfp'+R_+ \bfG''-R_- \bfG'-\bfq^J_n} e^{i ( R_+ \bfG'' \cdot \bft^{(1)}_\a - R_- \bfG' \cdot \bft^{(2)}_\b )}\, .
\eeq
\end{widetext}
In the above form, we make a series of approximations. Firstly, we expand around the momenta in the vicinity of the $K$-point in both layers
\beq
\bfp = R_+ K + \d \bfp \, , \qquad \bfp' = R_- K + \d \bfp' \, .
\eeq
As $J$ decays fast in the momentum space away from the rotated layer $K$-points $R_+ K$, $R_+ C_3 K$, $R_+ C_3^2 K$, we keep only $\bfG_1' = 0$, $\bfG_2' = C_3 K - K$, and $\bfG_3' = C_3^2 K - K$. For sufficiently small twist angles, the delta function imposes alike $\bfG_1'' = 0$, $\bfG_2'' = C_3 K - K$, and $\bfG_3'' = C_3^2 K - K$. Since we assume $\bfq^J_n \ll \bfG'$, we can let 
\beq
J^{(12)}_{\bfp' +R_+ \bfG' + \bfq_n^J} \simeq J^{(12)}_{\bfp' +R_+ \bfG'} \simeq \Omega J'\d_{\a \b} + \Omega J''(1-\d_{\a \b}) \, ,
\eeq
where we used the requirements imposed by a $C_3$ symmetry. We further introduce the moir\'e reciprocal lattice vectors that assume constant value in the small angle approximation
\beq
\bal
R_+ \bfG'_1 - R_- \bfG''_1 & = 0 \, , \\
R_+ \bfG'_2 - R_- \bfG''_2 & \equiv \bfG_{1}^{\rm M}\, , \\ 
R_+ \bfG'_3 - R_- \bfG''_3 & \equiv \bfG_{1}^{\rm M}+ \bfG_{2}^{\rm M} \, . \\ 
\eal
\eeq
The final form of this interlayer superexchange is then
\begin{widetext}
\beq
\bal
U^{(12)}_{\delta \bfp \delta \bfp'} & = -\f{1}{4} \sum_n
\begin{pmatrix}
    \chi_{AA}  J' & \chi_{AB}J'' \\
    \chi_{AB} J'' & \chi_{BB} J'
\end{pmatrix} \d_{\delta \bfp + \bfd_1 -\bfq^J_n, \delta \bfp'}  -\f{1}{4} \sum_n
\begin{pmatrix}
    \chi_{AA} J' & \w\chi_{AB} J'' \\
    \w^{-1} \chi_{AB} J'' & \chi_{BB} J'
\end{pmatrix} \d_{\delta \bfp + \bfd_2 -\bfq^J_n, \delta \bfp'} \\
& -\f{1}{4}  \sum_n
\begin{pmatrix}
    \chi_{AA} J' & \w^{-1}\chi_{AB} J'' \\
    \w \chi_{AB} J'' & \chi_{BB} J'
\end{pmatrix} \d_{\delta \bfp+\bfd_3-\bfq^J_n, \delta \bfp'}\, ,
\eal
\eeq
where $\omega = e^{2 \pi i/3}$. If we set $\bfq^J_n = \bfd_n$, unlike in the classic BM model, the next nearest neighbors on the moir\'e lattice can also hybridize as shown with the long arrows on the right panel of \autoref{hexplots}. The hybridization at zero momentum also becomes possible.

 We now discuss the Fourier transform of the $X$ term in \eqref{termsq}. It is convenient to perform this calculation separately for the $\bfq^J_n = 0$ and $\bfq^J_n = \bfd_n$ cases. We begin with the first case:
\beq
\bal
    \frac{\chi_{\a\b}^2}{4} \sum_{ij} J(|i-j|) & = \frac{\chi_{\a\b}^2}{4 \Omega N} \sum_{\bfR^{(1)} \bfR^{(2)} \bfG \bfq} J_{\bfG+\bfq} e^{i (\bfG+\bfq) \cdot (\bfR^{(1)} - \bfR^{(2)}+\bft^{(1)} - \bft^{(2)})} \\ &= \frac{N \chi_{\a\b}^2}{4 \Omega}\sum_{\bfG^{(1)} \bfG^{(2)} \bfG \bfq}J_{\bfG+\bfq} \d_{\bfG+\bfq,\bfG^{(1)}} \d_{\bfG+\bfq,\bfG^{(2)}} e^{i \bfG^{(1)} \cdot (\bft^{(1)} - \bft^{(2)})}\, .
    \eal
\eeq
At non-zero twist, the only matching reciprocal lattice vectors are $\bfG^{(1)}=0$ and $\bfG^{(2)}=0$, so we obtain
\beq
\frac{N \chi_{\a\b}^2}{4 \Omega}\sum_{\bfG \bfq}J_{\bfG+\bfq} \d_{\bfG+\bfq,0} = \frac{N \chi_{\a \b}^2}{4} J_0\, ,
\eeq
where we introduced $J_{\bfk=0} = \Omega J_0$. An analogous calculation for the $\bfq^J_n = \bfd_n$ case proceeds as follows:
\beq
\bal
    \frac{\chi^2_{\a\b}}{4}  \sum_{ijnn'} J(|i-j|) & e^{-i (\bfd_{n'} - \bfd_{n}) \cdot j}  = \frac{\chi_{\a\b}^2}{4 \Omega N} \sum_{\bfR^{(1)} \bfR^{(2)} \bfG \bfq \, n n'} J_{\bfG+\bfq} e^{i (\bfG+\bfq) \cdot (\bfR^{(1)} - \bfR^{(2)}+\bft^{(1)} - \bft^{(2)})} e^{-i (\bfd_{n'} - \bfd_{n}) \cdot (\bfR^{(2)} + \bft^{(2)})} \\ &= \frac{N \chi_{\a\b}^2}{4 \Omega}\sum_{\bfG^{(1)} \bfG^{(2)} \bfG \bfq\, n n'}J_{\bfG+\bfq} \d_{\bfG+\bfq,\bfG^{(1)}} \d_{\bfG+\bfq+\bfd_{n'} - \bfd_{n},\bfG^{(2)}} e^{i \bfG^{(1)} \cdot (\bft^{(1)} - \bft^{(2)})} e^{-i (\bfd_{n'} - \bfd_{n}) \cdot \bft^{(2)}}\, .
    \eal
\eeq
We further sum over $\bfG \bfq$ obtaining

\beq
\frac{N \chi_{\a\b}^2}{4 \Omega}\sum_{\bfG^{(1)} \bfG^{(2)}\, n n'}J_{\bfG^{(1)}} \d_{\bfG^{(1)},\bfG^{(2)}+\bfd_n - \bfd_{n'}} e^{i \bfG^{(1)} \cdot (\bft^{(1)} - \bft^{(2)})} e^{i (\bfd_n - \bfd_{n'}) \cdot  \bft^{(2)}}\, .
\label{tablesum}
\eeq
\end{widetext}
Since $\bfG^{(1)}$ and $\bfG^{(2)}$ differ only by a unit reciprocal lattice vector, the momentum conservation allows for the possibilities listed in \autoref{chi2tab}.

\begin{table}[h!tb]
\centering

\begin{tabular}{|c|c|c|c|}
\hline $\bfG^{(1)}$ & $\bfG^{(2)}$ & $\bfG^{(1)}-\bfG^{(2)}$ & $\left(n, n^{\prime}\right)$ \\
\hline\hline 0 & 0 & 0 & (1,1),(2,2),(3,3) \\
\hline $\bfG_1^{(1)}$ & $\bfG_1^{(2)}$ & $\bfG_{1}^{ \rm M}$ & (2,1) \\
\hline $\bfG_2^{(1)}$ & $\bfG_2^{(2)}$ & $\bfG_{2}^{ \rm M}$ & (3,1) \\
\hline $\bfG_1^{(1)}-\bfG_2^{(1)}$ & $\bfG_1^{(2)}-\bfG_2^{(2)}$ & $\bfG_{1}^{ \rm M}-\bfG_{2}^{ \rm M}$ & (2,3) \\
\hline $-\bfG_1^{(1)}$ & $-\bfG_1^{(2)}$ & $-\bfG_{1}^{ \rm M}$ & (1,2) \\
\hline$-\bfG_2^{(1)}$ & $-\bfG_2^{(2)}$ & $-\bfG_{2}^{ \rm M}$ & (1,3) \\
\hline$-\bfG_1^{(1)}+\bfG_2^{(1)}$ & $-\bfG_1^{(2)}+\bfG_2^{(2)}$ & $-\bfG_{1}^{\rm M}+\bfG_{2}^{ \rm M}$ & (3,2) \\
\hline
\end{tabular}
\caption{Combinations of summation variables that yield non-zero results in the sum \eqref{tablesum}}
\label{chi2tab}
\end{table}

Using the delta function, we then obtain, depending on the sublattice indices:\\

--- $\chi_{AA}$ terms ($\bft^{(1)}=0, \bft^{(2)} = 0$) or $\chi_{BB}$ terms ($\bft^{(1)}\neq 0, \bft^{(2)} \neq 0$):
\beq
 \frac{3 N \chi_{\a\a}^2 J_0}{4} +  \frac{6 N \chi_{\a\a}^2 \bar J}{4} \, ,
\eeq
where $ J_{\bfk = \bfG_1} = J_{\bfk = \bfG_2} = \ldots \equiv  \Omega \bar J$.\\ --- $\chi_{AB}$ terms ($\bft^{(1)} = 0, \bft^{(2)} \neq 0$) or $\chi_{BA}$ terms ($\bft^{(1)} \neq 0, \bft^{(2)} = 0$):
\begin{widetext}
\beq
\bal
 \frac{3 N \chi_{AB}^2 J_0}{4} + \frac{N \chi_{AB}^2 \bar J}{4} & \( e^{\mp i \bfG_{1} \cdot \bft} + e^{\mp i \bfG_{2} \cdot \bft} + e^{\mp i (\bfG_{1}+\bfG_{2}) \cdot \bft}  + {\rm c.c.}\) \\
& =  \frac{3 N \chi_{AB}^2 J_0}{4} + \frac{N \chi_{AB}^2 \bar J}{4} \( e^{\mp 2 \pi i/3} + e^{\pm 2 \pi i/3} + 1  + {\rm c.c.}\) = \frac{3 N \chi_{AB}^2 J_0}{4} \, .
\eal
\eeq
\end{widetext}
As the last (${\cal N}$) term in \eqref{termsq} can be treated analogously, this concludes the derivation of the continuum model of the twisted hexagonal Mott insulator. The complete versions of the model with $\bfq^J_n = 0$ and $\bfq^J_n = \bfd_n$ together with the mean-field equations are summarized below.


\subsection{Uniform $\chi$ case}

We write the mean-field Hamiltonian for the twisted hexagonal bilayer as a sum of the intralayer and the interlayer parts
\beq
\hat H_{\rm u} = \hat H^{\rm intra}+C^{\rm intra}+\hat H^{\rm inter}_{\rm u}+C^{\rm inter}_{\rm u}\, .
\label{huapp}
\eeq

We further perform on $\hat H_{\rm u}$ the gauge transformation with a unitary matrix
\begin{widetext}
\begin{equation}
    {\mathcal{U}}_{\d \bfp \d \bfp'} = \begin{pmatrix}
        \delta_{\d \bfp, \d \bfp' - \bfd_1/2} & 0 & 0 & 0 \\
        0 & \delta_{\d\bfp, \d\bfp' - \bfd_1/2}  & 0 & 0 \\
        0 & 0 &\delta_{\d\bfp, \d\bfp' + \bfd_1/2} & 0 \\
        0 & 0 & 0 &\delta_{\d\bfp, \d\bfp' + \bfd_1/2} 
    \end{pmatrix} \, 
\end{equation}
simultaneously with a shift of variables $\delta \bfp \to \delta \bfp - \bfd_1/2$, $\delta \bfp' \to \delta \bfp' + \bfd_1/2$. Just like in the Bistritzer-Macdonald model, in this basis, only the states in different layers with momenta that differ by $\bfd_n$ can hybridize. This makes it convenient to parametrize
\begin{equation}
    \delta \bfp = \bfQ+\bfk \, , \qquad \delta \bfp' = \bfQ'+\bfk \, ,
\end{equation}
 where $\bfQ$ and $\bfQ'$ belong to the different sublattices of the honeycomb lattice built on vectors $\bfd_n$. 

After this procedure, the intralayer term takes the form
\beq
\(\hat H^{\rm intra}\)_{\bfQ \bfQ'} = \d_{\bfQ \bfQ'}  \sum_{\bfk} \hat f^\dagger_{\bfk} H^{\rm intra}(\bfk) \hat f_{\bfk} \, ,
\eeq
where
\beq
H^{\rm intra}(\bfk) = \begin{pmatrix}
\l - \mu_0 & - \( \d t + \f{\chi J}{4}\)K_\bfk^{(1)}  & 0 & 0 \\
- \( \d t + \f{\chi J}{4}\) K_\bfk^{(1)}  & \l - \mu_0 & 0 & 0 \\
0 & 0 & \l - \mu_0 & - \( \d t + \f{\chi J}{4}\) K_\bfk^{(2)}  \\
 0 & 0 & - \( \d t + \f{\chi J}{4}\) K_\bfk^{(2)}  & \l - \mu_0
\end{pmatrix} \, ,
\eeq
and the rotated hopping matrices are
\beq
 K_\bfk^{(l)} = v_F \lb R(\pm \theta/2) (\bfk+\bfQ)\cdot(\hat \s_x, \hat \s_y) \rb\, , \qquad  v_F = \f{3}{2} a t \, .
\eeq
The intralayer constant term is
\beq
C^{\rm intra} = \frac{3}{2}N J \chi^2 + \frac{N}{2}(1-\d)(3 \d J - 8 \l)\, .
\eeq
The interlayer Hamiltonian can be cast as
    \beq
(\hat H^{{\rm inter}}_{\rm u})_{\bfQ \bfQ'} =\sum_{\bfk \l} \hat f^\dagger_{\bfk}\( H^{{\rm inter}}_{1} \d_{\bfQ' - \bfQ,- \l \bfd_1}  + H^{{\rm inter}}_{2} \d_{\bfQ' - \bfQ,- \l \bfd_2}  + H^{{\rm inter}}_{3} \d_{\bfQ' - \bfQ, - \l \bfd_3}  \) \hat f_{\bfk}\, ,
\eeq
where the $\lambda$-summation is over two values: $\l=+1$ corresponds to the hopping from layer 2 to layer 1, and $\l=-1$ in the opposite direction. We further define
\beq
H^{\rm inter}_{i} = 
- \begin{pmatrix}
0 & 0  & \d t' +\f{\chi_{\rm AA}J' }{4}&  \w^{i-1}\(\d t'' +\f{\chi_{\rm AB}J''}{4}\) \\
0  & 0 & \w^{-(i-1)}\( \d t'' +\f{\chi_{\rm AB}J'' }{4} \) & \d t' +\f{\chi_{\rm BB}J' }{4} \\
0 & 0 & 0 & 0  \\
 0 & 0 & 0  & 0
\end{pmatrix} \d_{\l,1} + \({\rm h. c.}\) \d_{\l,-1} \, .
\eeq
Lastly, the constant term arising from the interlayer tunneling and superexchange reads
\beq
C^{\rm inter}_{\rm u} = \frac{N}{4} \(  \chi_{\rm AA} ^2 +  \chi_{\rm BB}^2 \) J_0'  +   \frac{N}{2} \chi_{\rm AB}^2 J_0''  +  \f{N }{2}\d(1-\d) \( J_0'  +J_0'' \) \, .
\eeq
The mean-field equations for $\chi$, $\chi_{\rm AA}$, $\chi_{\rm BB}$, $\chi_{\rm AB}$,  $\l$, $\d$ obtained by minimizing \eqref{huapp} are as follows:
\begin{eqnarray}
   \chi & =& \frac{1}{6} \sum_{\bfQ,\, n<E_F} \int \f{d^2 \bfk}{K} \braket{n \bfk \bfQ|\hat H^{\rm intra}|n \bfk \bfQ} \, , \nonumber \\
\chi_{\rm AA(BB)} & =& \f{J'}{J_0'} \sum_{\bfQ,\, n<E_F} \int \f{d^2 \bfk}{K} \braket{n \bfk \bfQ|\hat H^{\rm AA(BB)}_{\rm u}|n \bfk \bfQ} \, , \nonumber \\
\chi_{\rm AB} & =&\f{J''}{2J_0''} \sum_{\bfQ,\, n<E_F} \int \f{d^2 \bfk}{K} \braket{n \bfk \bfQ|\hat H^{\rm AB}_{\rm u}|n \bfk \bfQ} \, , \\
\d & =& 1 - \half \sum_{\bfQ, n<E_F} \int \f{d^2 \bfk}{K}\braket{n \bfk \bfQ | \hat H_{\rm u} | n \bfk \bfQ} \, ,\nonumber \\
\l & =& 3 t \chi + \frac{ t'}{2} \frac{J_0'}{J'}\(\chi_{\rm AA}+\chi_{\rm BB}\) + \frac{2 J_0''}{J''} \f{t''}{2} \chi_{\rm AB} \nonumber  -  \frac{1}{8}(1-2 \d)(3 J +  J_0' + J_0'') \nonumber \, , 
\label{sc1}
\end{eqnarray}

where

\beq
\(\hat H^{\rm intra}\)_{\bfQ \bfQ'} = \d_{\bfQ \bfQ'} \sum_{\bfk} \hat f^\dagger_{\bfk}
\begin{pmatrix}
0 & K_\bfk^{(1)}  & 0 & 0 \\
 K_\bfk^{(1)}  & 0 & 0 & 0 \\
0 & 0 & 0 &  K_\bfk^{(2)}  \\
 0 & 0 &  K_\bfk^{(2)}  & 0
\end{pmatrix} f^\dagger_{\bfk} \, ,
\eeq
and 
\beq
(\hat H^{\rm AA(BB,AB)}_{\rm u})_{\bfQ \bfQ'} =\sum_{\bfk \l} \hat f^\dagger_{\bfk}\( H^{\rm AA(BB,AB)}_{1} \d_{\bfQ' - \bfQ, - \l \bfd_1}  + H^{\rm AA(BB,AB)}_{2} \d_{\bfQ' - \bfQ,- \l \bfd_2}  + H^{\rm AA(BB,AB)}_{3} \d_{\bfQ' - \bfQ, - \l \bfd_3}  \) \hat f_{\bfk}\, ,
\eeq
where, finally, 
\begin{equation}
\begin{aligned}
    H^{\rm AA}_{i} = 
\begin{pmatrix}
    0 & 0 & 1 & 0\\
    0 & 0 & 0 & 0 \\
    0 & 0 & 0 & 0\\
    0 & 0 & 0 & 0
\end{pmatrix}\d_{\l,1} + \({\rm h. c.}\)  \d_{\l,-1}\, , ~~ H^{\rm BB}_{i} = & 
\begin{pmatrix}
    0 & 0 & 0 & 0\\
    0 & 0 & 0 & 1 \\
    0 & 0 & 0 & 0\\
    0 & 0 & 0 & 0
\end{pmatrix}\d_{\l,1} + \({\rm h. c.}\) \d_{\l,-1}\, ,\\
~~ H^{\rm AB}_{i} = 
\begin{pmatrix}
    0 & 0 & 0 & \omega^{i-1}\\
    0 & 0 & \omega^{-(i-1)} & 0 \\
    0 & 0 & 0 & 0\\
    0 & 0 & 0 & 0
\end{pmatrix}& \d_{\l,1} + \({\rm h. c.}\) \d_{\l,-1}\, .\end{aligned}
\end{equation}

\subsection{Spatially-dependent $\chi$ case}
We keep the same gauge as in the uniform case and present the Hamiltonian as below:
\begin{equation}
    \hat H_{\rm nu} = \hat H^{\rm intra} +C^{\rm intra} + \hat H^{\rm inter}_{\rm nu} +C^{\rm inter}_{\rm nu} \, .
    \label{hnuapp}
\end{equation}
The third term in this expression is
\beq
(H^{{\rm inter}}_{\rm nu})_{\bfQ \bfQ'}  = \sum_{\bfk \l n} \hat f^\dagger_{\bfk}\left( H^{\rm inter}_{1,J} \d_{\bfQ' - \bfQ,~ -\l(\bfd_1- \bfd_n)} \right.   \left.+ H^{\rm inter}_{2,J} \d_{\bfQ' - \bfQ,~- \l (\bfd_2-\bfd_n)} \right. \left.+ H^{\rm inter}_{3,J} \d_{\bfQ' - \bfQ, ~- \l (\bfd_3-\bfd_n)}  \right) \hat f_{\bfk} + \hat H_{{\rm u},t'}^{\rm inter}\, ,
\label{nonunihamapp}
\eeq 
\end{widetext}
where we defined
\begin{equation}
    H^{\rm inter}_{i,J} = \left. H^{\rm inter}_{i} \right|_{\delta \to 0}\, , \qquad \hat H_{{\rm u},t'}^{\rm inter} = \left. \hat H_{{\rm u}}^{\rm inter} \right|_{J',J'' \to 0}\, .
\end{equation}
The crucial novelty introduced by the presence of extra $\bfd_n$ in \eqref{nonunihamapp} is the possibility of hybridization between two Dirac cones from different layers at the momenta $\bfd_n - \bfd_m$. This necessitates promoting $\bfQ$, $\bfQ'$ to the coordinates of the vertices belonging to the \textit{triangular} lattice spanned by vectors $\bfd_n$, which are the reciprocal lattice vectors in this model. This change makes the area of the Brillouin zone three times smaller than in the uniform order parameter case solved on a honeycomb lattice spanned by the same vectors. Consequently, the size of moir\'e unit cell in the real space becomes three times larger. Lastly, the constant coming from the interlayer hopping is $C^{\rm inter}_{\rm nu} =\frac{3 N}{4}(J_0'+2 \bar J') \(  \chi_{\rm AA} ^2 +  \chi_{\rm BB}^2 \) +   \frac{3N}{2} \chi_{\rm AB}^2 J_0''  +  \f{N }{2}\d(1-\d) \( J_0'  +J_0'' \) $. The mean-field equations for $\chi$, $\chi_{\rm AA}$, $\chi_{\rm BB}$, $\chi_{\rm AB}$,  $\l$, $\d$ are only slightly modified compared to the case with the uniform order parameter \eqref{sc1}
\begin{widetext}
\beq
\bal
\chi & = \frac{1}{6} \sum_{\bfQ,\, n<E_F} \int \f{d^2 \bfk}{K} \braket{n \bfk \bfQ|\hat H^{\rm intra}|n \bfk \bfG_M} \, ,\\
\chi_{\rm AA(BB)} & = \f{J'}{3(J_0'+2 \bar J')} \sum_{\bfQ,\, n<E_F} \int \f{d^2 \bfk}{K} \braket{n \bfk \bfQ|\hat H^{\rm AA(BB)}_J|n \bfk \bfQ}\\
\chi_{\rm AB} & =\f{J''}{6J_0''} \sum_{\bfQ,\, n<E_F} \int \f{d^2 \bfk}{K} \braket{n \bfk \bfQ|\hat H^{\rm AB}_J|n \bfk \bfQ}\, ,\\
\d & = 1 - \half \sum_{\bfQ, n<E_F} \int \f{d^2 \bfk}{K}\braket{n \bfk \bfG_M | \hat H_{\rm nu} | n \bfk \bfQ} \, ,\\
\l & = 3 t \chi + \frac{ t'}{2} \(I^t_{AA}+I^t_{BB}\) + \f{t''}{2} I^t_{AB} - \frac{1}{8}(1-2 \d)(3 J +  J_0' + J_0'')\, ,
\eal
\label{sc2}
\eeq
with
\begin{equation}
\begin{aligned}
(\hat H^{\rm AA(BB,AB)}_J)_{\bfQ \bfQ'} =\sum_{\bfk \l} \hat f^\dagger_{\bfk}\left( H^{\rm AA(BB,AB)}_{1} \d_{\bfQ' - \bfQ,~ -\l (\bfd_1 - \bfd_n)}  + H^{\rm AA(BB,AB)}_{2} \right. &  \d_{\bfQ' - \bfQ,~-\l (\bfd_2 - \bfd_n)}  \\  + & \left. H^{\rm AA(BB,AB)}_{3} \d_{\bfQ' - \bfQ,~-\l (\bfd_3 - \bfd_n)}  \right) \hat f_{\bfk}\, .
\end{aligned}
\end{equation}
\end{widetext}
\section{Symmetries of the non-uniform $\chi$ model}

\label{appc}

\begin{figure}[t!] 
    \centering
        \centering
        \includegraphics[width=0.95\linewidth]{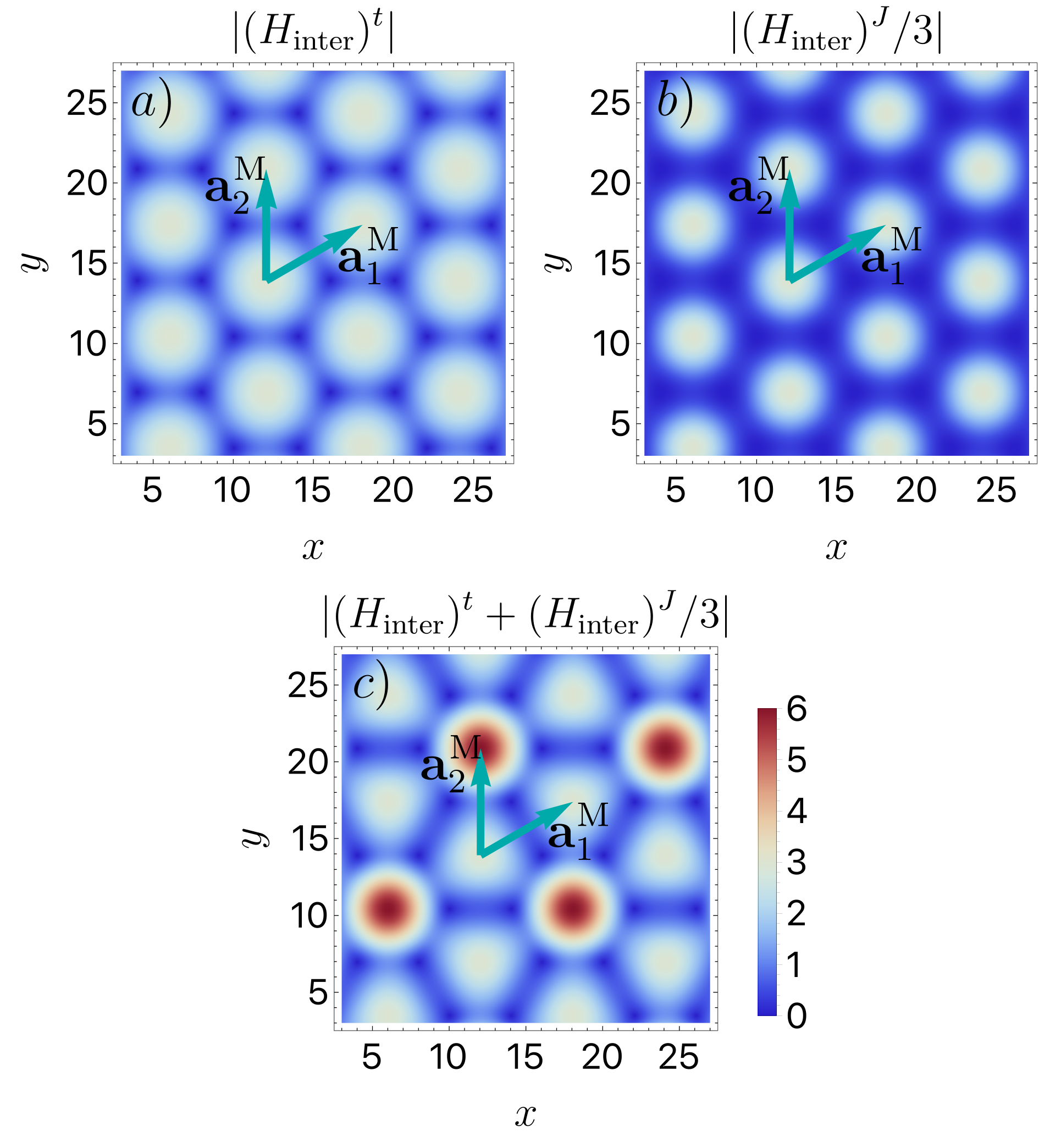}
    \caption{The textures in the real space corresponding to the terms $|(H_{\rm inter})^{t}|$ ($a$), $|(H_{\rm inter})^{J}/3|$ ($b$) and their interference pattern $|(H_{\rm inter})^{t} + (H_{\rm inter})^{J}/3|$ ($c$). The moir\'e lattice vectors $\bfa_{1,2}^{\rm M}$ are indicated with blue arrows. As one can infer from the figures, the first two terms are moir\'e-periodic, whereas the latter breaks moir\'e translations.}
    \label{textures}
\end{figure}

The symmetries of the BM model are well-known \cite{bernevig}. In our model, on the other hand, we leave the possibility for the mismatch in the momentum dependence between the terms responsible for the interlayer tunneling (denoted with $t$) and the interlayer superexchange (denoted with $J$). Both types of terms can be cast in the form
\begin{widetext}
\begin{equation}
\begin{gathered}
    (H_{\rm inter})^{t(J)}_{\delta\bfp \delta{\bfp'}}  = \sum_{\bfk \l n m} \hat f^\dagger_{\delta\bfp} H_{{\rm inter},m\l}^{t(J)}\hat f_{\delta\bfp'} \d_{\delta\bfp' - \delta\bfp,~- \l (\bfd_m- \bfq_{ n}^{t(J)})}\, , \\
H_{{\rm inter}, m \l}^t = 
- \begin{pmatrix}
0 &  T^t_m \\
0  & 0
\end{pmatrix} \d_{\l,1} + \({\rm h. c.}\) \d_{\l,-1} \, , \qquad 
H_{{\rm inter}, m \l}^J = 
- \begin{pmatrix}
0 &  T^J_m \\
0  & 0
\end{pmatrix} \d_{\l,1} + \({\rm h. c.}\) \d_{\l,-1} \, , \\
    T_m^t = \begin{pmatrix}
         \d t'&  \w^{m-1} \d t'' \\
         \w^{-(m-1)} \d t''  & \d t' 
    \end{pmatrix} \, , \qquad
    T_m^J = \begin{pmatrix}
        \f{\chi_{AA}J' }{4}&  \w^{m-1}\f{\chi_{AB}J''}{4} \\
        \w^{-(m-1)} \f{\chi_{AB}J'' }{4} &  +\f{\chi_{BB}J' }{4} 
    \end{pmatrix} \, .
    \end{gathered}
\end{equation}
\end{widetext}
Furthermore, both interlayer terms can be diagonalized by performing the Fourier transform over $\delta \bfp$ and $\delta \bfp'$, which yields
\begin{equation}
    (H_{\rm inter})^{t(J)}  = \sum_{j \l n m} \hat f^\dagger_j H_{{\rm inter},m\l}^{t(J)}\hat f_j e^{i(\bfd_m- \bfq_{ n}^{t(J)}) \cdot \bfj} \, .
\end{equation}
Let us consider how the moir\'e translation symmetry acts on the sum $(H_{\rm inter})^t + (H_{\rm inter})^J$. The interlayer hopping term in our model has a trivial momentum dependence, i.e. $\bfq^t_n = 0$. Therefore, if the interlayer superexchange term has no momentum dependence as well ($\bfq^J_n = 0$), the translational symmetry of the Hamiltonian that includes the sum of $(H_{\rm inter})^{t}$ and $(H_{\rm inter})^{J}$ is identical to that in the Bistritzer-Macdonald model. Hence, moir\'e translations are a good symmetry in this case.

Now suppose $\bfq_n^J = \bfd_n$. Consider performing a coordinate transformation by shifting $\bfj$ by one of the moir\'e lattice vectors $\bfa_{1} = 3a/(2 \theta)(1,\, 1/\sqrt{3})$. Since $e^{\bfa_{1} \cdot \bfd_n} = \omega$, and $e^{\bfa_{1} \cdot (\bfd_{n} - \bfd_{m})} = 1$, the term $(H_{\rm inter})^{t}$ acquires the additional phase equal to $\omega$ under moir\'e translation, whereas the interlayer superexchange $(H_{\rm inter})^{J}$ stays invariant. If the superexchange term was absent, one could have performed a gauge transformation with the matrix
\begin{equation}
    {\cal M} = \begin{pmatrix}
        \omega^{1/2} & 0 & 0 & 0\\
        0 & \omega^{1/2} & 0 & 0\\
        0 & 0 & \omega^{-1/2} & 0\\
        0 & 0 & 0 & \omega^{-1/2}
    \end{pmatrix} \, ,
\end{equation}
which would have restored $(H_{\rm inter})^{t}$ to the original form. If, on the other hand, both superexchange and the hopping terms are present, this gauge transformation cannot bring their sum to the original form. Therefore, having $\bfq_n^J = \bfd_n$ simultaneously with $\delta \neq 0$ lowers the translation symmetry. The new translation vectors $\bfdd_{1} = 2a/(2 \theta)(1,\,1/\sqrt{3})$, $\bfdd_2 = (3a/\theta,0)$ are the reciprocal lattice vectors with respect to $\bfd_n$, since in this case $e^{\bfdd_{i} \cdot \bfd_n} = e^{\bfdd_i \cdot \bfG_{n}^{\rm M}} = 1$, and such translation does not produce any phases. This argument can be visualized by considering the textures in the real space produced by the terms $|(H_{\rm inter})^{t}|$, $|(H_{\rm inter})^{J}/3|$, and $|(H_{\rm inter})^{t} + (H_{\rm inter})^{J}/3|$ with all the constant prefactors in these terms such as $\delta t'$ or $\chi_{\rm AA} J'/4$ set to unity. As one can see from \autoref{textures}, the former two terms produce a moir\'e-periodic pattern in real space, whereas the latter breaks moir\'e translations.

Besides the moir\'e translations, in our model, the $C_2$ symmetry can also be broken if $\chi_{\rm AA} \neq \chi_{\rm BB}$. In order for $(H_{\rm inter})^{J}$ to be invariant under $C_2$, one requires \cite{bernevig} 
\beq
\sigma_x T_1^J \sigma_x = T_1^J \qquad \sigma_x T_2^J \sigma_x = T_3^J \, .
\eeq
Since the hopping matrix interchanges both the diagonal and off-diagonal entries and the off-diagonal terms are self-consistently found to be equal across all phases, the case $\chi_{\rm AA} = \chi_{\rm BB}$ preserves this symmetry and $\chi_{\rm AA} \neq \chi_{\rm BB}$ breaks.

\section{Comparison with the untwisted case and twist angle dependence}

\label{appd}



\begin{figure}[t!]
\centering
\includegraphics[width=0.8\linewidth]{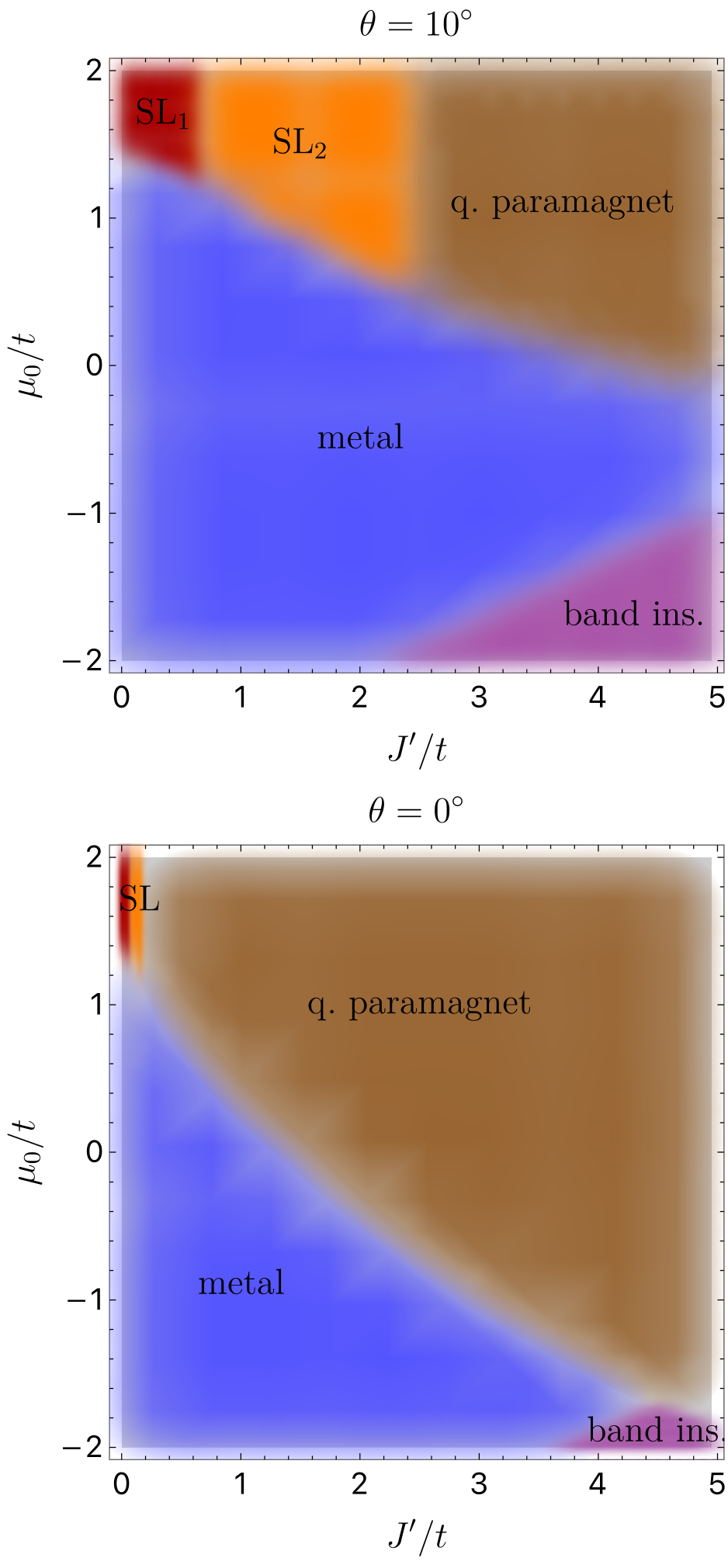}
\caption{The comparison of the phase diagram between the $\theta=10^\circ$ (top) and the untwisted case (on the bottom). The band insulator phase is characterized by $\delta=1$, i.e. no spinons are present in the system. The parameters used to obtain the diagrams are matched according to the procedure outlined in \autoref{appd}. The variables not fixed by matching are the same as in \autoref{diag}. We note that for the values of parameters used, no excitonic insulator phase is found due to the large ratio $J_0'/J'$ required for matching.}
\label{twuntw}
\end{figure}
\begin{figure}[t!]
\centering
\includegraphics[width=1\linewidth]{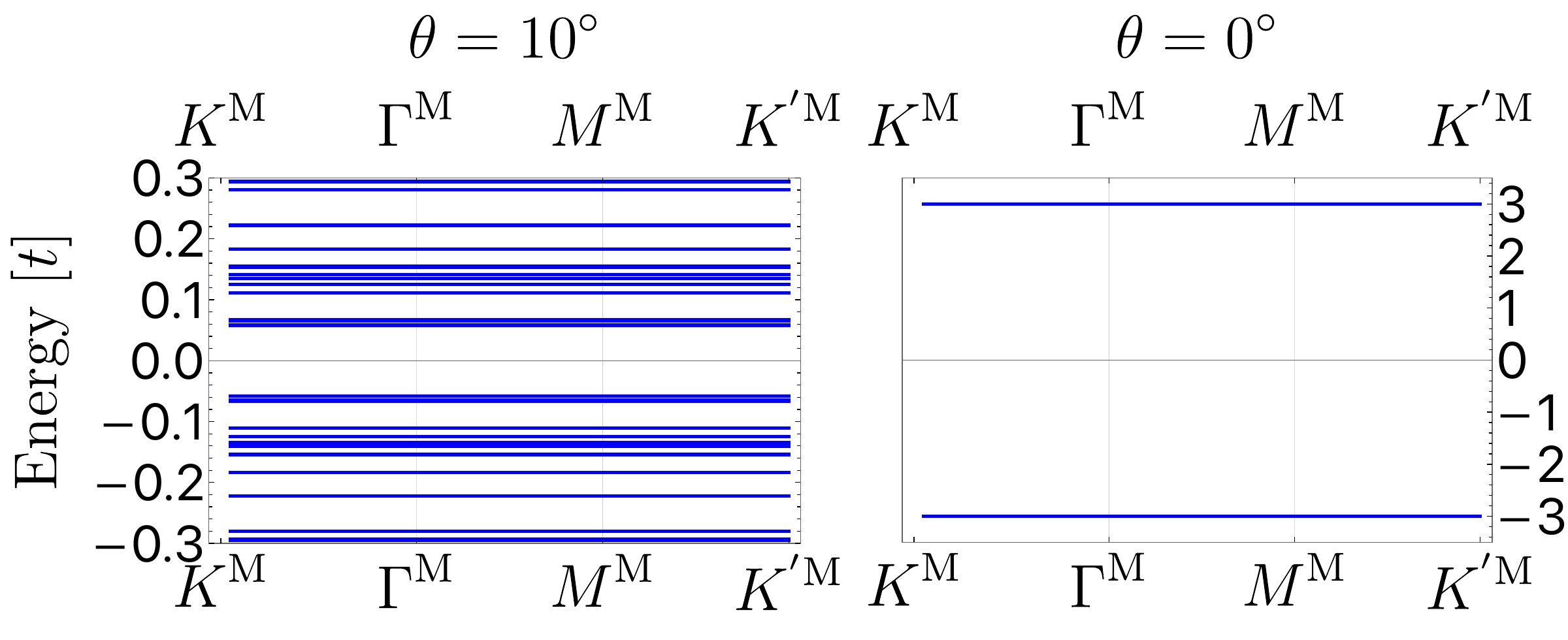}
\caption{Band structure in the quantum paramagnet case with $\theta=10^\circ$ (on the left) and $\theta=0^\circ$ on the right. The parameters used correspond to the $J' = 4 t$, $\mu = 1.5 t$ in the phase diagrams in \autoref{twuntw}.}
\label{qpbs}
\end{figure}

In this appendix, we compare the model of the untwisted correlated bilayer introduced in \autoref{appa} and the model describing the correlated twisted bilayer obtained in \autoref{appb}.

Let us consider as a starting point the untwisted case defined in the full hexagonal Brillouin zone (BZ). The matching onto the continuous model is performed by dividing the Brillouin zone into two Dirac cones with the conservation of the number of states
\beq
\int_{\rm BZ} \f{d^2 \bfk \, A}{(2 \pi)^2} = 2 \int_{\rm |\bfk|<R} \f{d^2 \bfk \, A}{(2 \pi)^2} \, .
\eeq
Therefore,
\beq
\f{8 \pi^2 a^2}{3 \sqrt{3}} = 2 \pi R^2 \to R = \sqrt{\frac{4 \pi}{3 \sqrt{3}}} a \, .
\eeq
We expect this truncation of the Brillouin zone to introduce some errors that we discuss further below.

The next step is to match the full untwisted Hamiltonian \autoref{haaapp} with both the uniform \autoref{huapp} and non-uniform cases \autoref{hnuapp}. The uniform model reduces to the untwisted case if, with $\theta = 0$, we simultaneously set $J_0' = 3 J'$.  We also need to use $\chi_{\rm AB} = 0$ and $J_0''=0$, $t''=0$ which ensures that no traces of AB superexchange remain. It can be further seen in the non-uniform case that up to the re-definition of $\chi_{\rm AA}$ and $\chi_{\rm BB}$, the Hamiltonian for the untwisted bilayer can be reproduced by setting $J_0' = \bar J' = 3 J'$, $\chi_{\rm AB} = 0$, $J_0''=0$.

We immediately note that the nature of the interlayer hybridization is completely different in the $\theta = 0$ and $\theta \simeq 0$ cases, and no smooth crossover exists between the two. In the former case, the gap may open only at neutrality, whereas in the latter, the minibands hybridize with each other, and many gaps are present at the same time. Therefore, whenever interlayer hybridization is important, we expect considerable differences between the twisted and untwisted cases even at small angles, whereas the properties of the less reliant on the existence of the gaps metallic phase should be relatively similar. 

This intuition is confirmed by the numerical simulations, as seen in \autoref{twuntw}. We observe that the quantum paramagnet phase in the twisted case occupies a much smaller fraction of the diagram due to the difference in the hybridization pattern: two remote flat bands occur in the $\theta = 0^\circ$ phase, whereas multiple flat bands in $\theta = 10^\circ$ phase, as shown in \autoref{qpbs}. The latter configuration is much less efficient in minimizing the interlayer part of the energy as the gap is effectively smaller. We also point out that for the same reason, the spin liquid phases are almost absent on the $\theta = 0^\circ$ diagram, whereas the twisted case features such phases.

\begin{figure}[t!]
\centering
\includegraphics[width=0.9\linewidth]{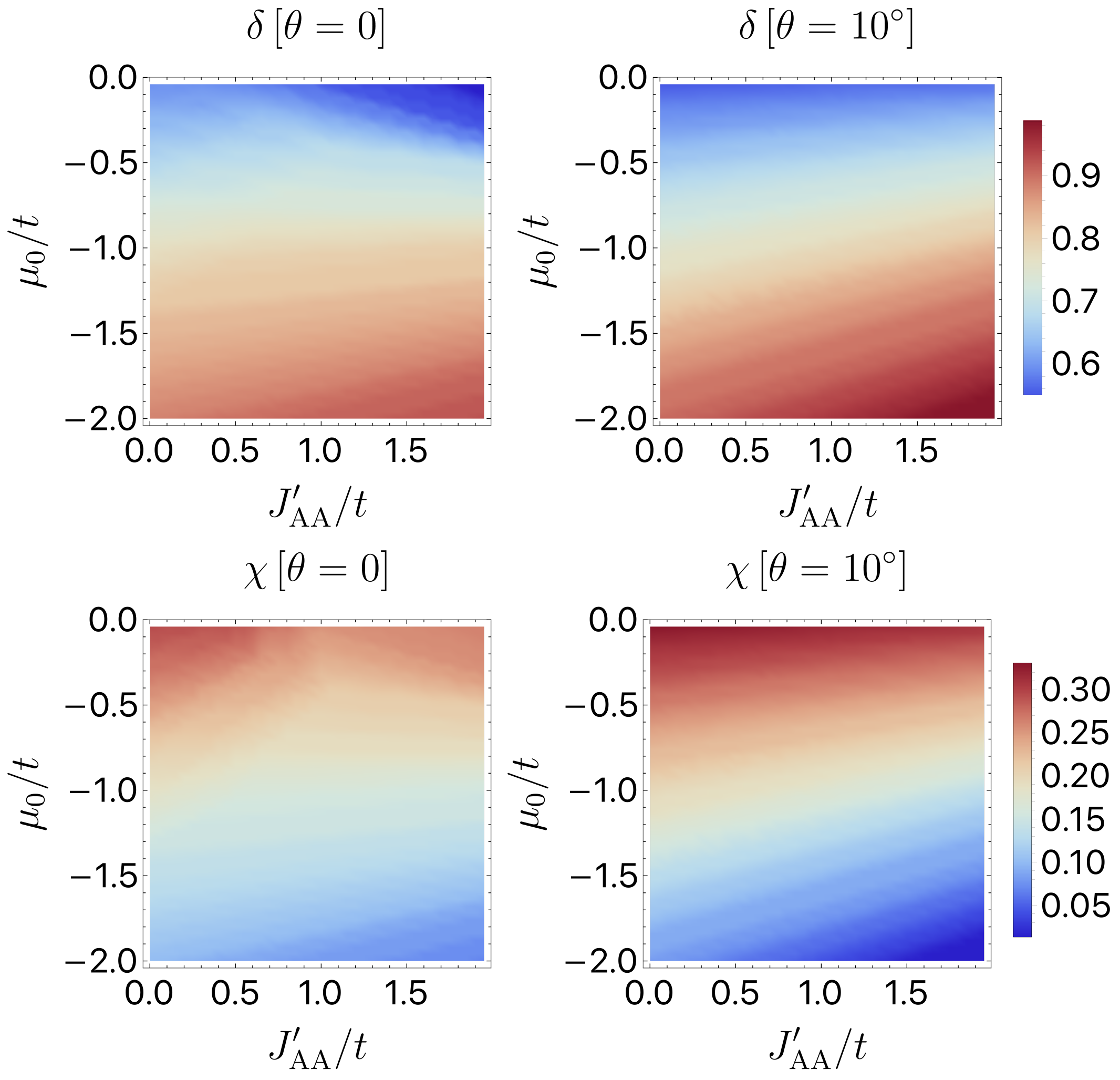}
\caption{Comparison of mean-field parameters $\delta$ and $\chi$ between the metallic phases in the twisted and untwisted cases in \autoref{twuntw}. The quantities agree well between $\theta=10^\circ$ and $\theta=0^\circ$ cases which justifies the matching procedure.}
\label{metcomp}
\end{figure}

As we expected, the metallic phases in both cases show very similar properties as follows from the \autoref{metcomp} in which the parameters $\delta$ and $\chi$ are compared between the twisted ($\theta = 10^\circ$) and the untwisted cases. From this, we conclude that the truncation of the Brillouin zone down to two circles surrounding the Dirac points does not qualitatively affect our results.

\begin{figure}[t!]
\centering
\includegraphics[width=0.7\textwidth]{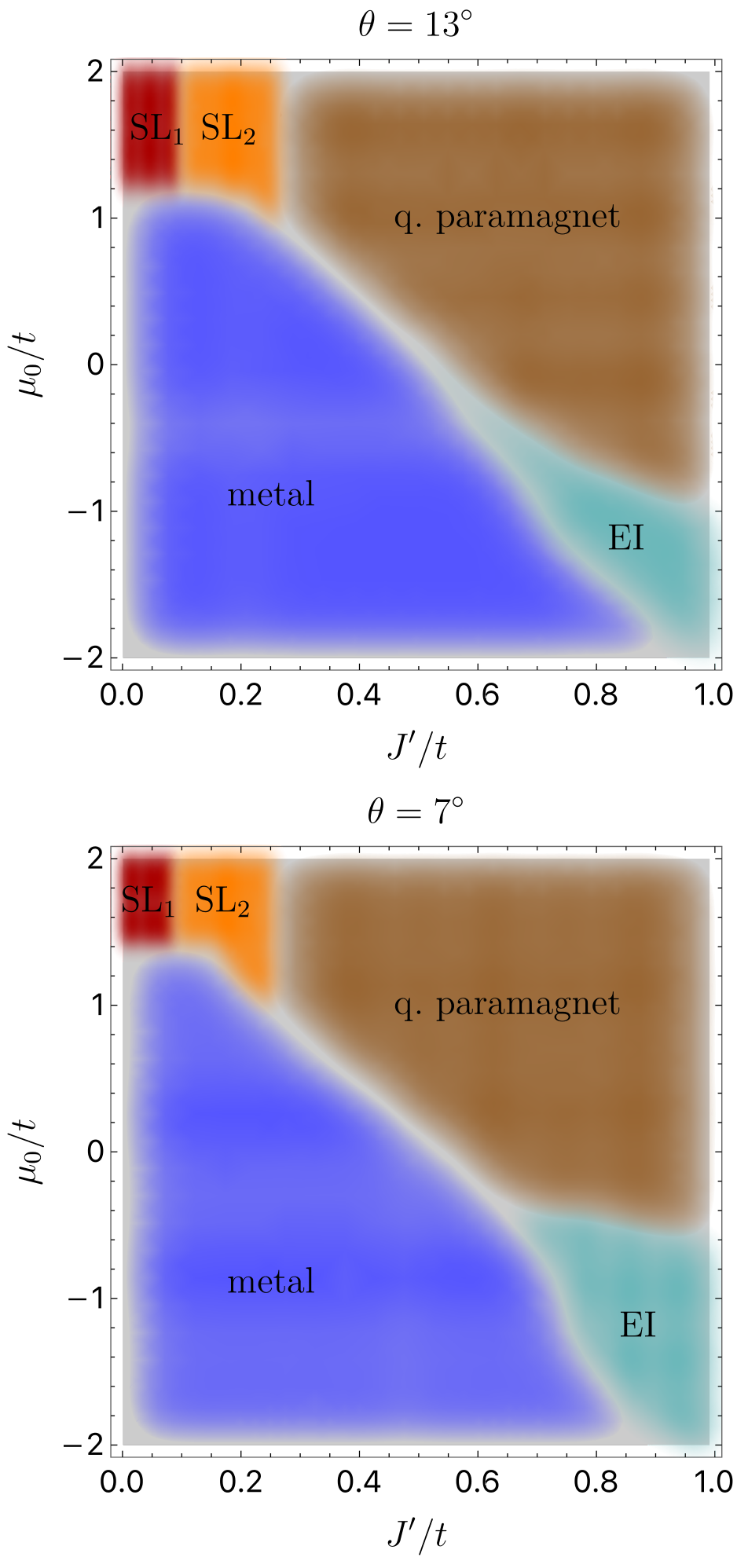}
\caption{Phase diagrams obtained for the same set of parameters as used in \autoref{diag} but for different twist angles: we set $\theta=13^\circ$ for the top figure and $\theta=7^\circ$ for the bottom.}
\label{twdep}
\end{figure}

 Lastly, we comment on the twist angle variation of the phase diagram: the phase diagrams for $\theta=13^\circ$, and $\theta = 7^\circ$ are shown in \autoref{twdep}. One would expect that the lower the twist angle is, the more there are minibands, and the opportunity to stabilize the gapped EI phase characterized by the small but non-zero $\delta$ is increasing. This trend is seen among the calculated phase diagrams presented in \autoref{twdep}: the EI phase is the only one essentially sensitive to the increase of the twist angle, and would eventually disappear as $\theta$ is increased. The reason is that at the large twist angle, only two Dirac cones within the cutoff radius $R$ remain, and the minibands are absent. Therefore, the phase diagram at large $\theta$ essentially becomes the one corresponding to the untwisted case.

\clearpage

\bibliography{literature}

\end{document}